\documentclass[twocolumn,superscriptaddress,aps,prc,floatfix]{revtex4}
\usepackage{graphicx}
\usepackage{dcolumn}
\usepackage{bm}
\usepackage{longtable}
\usepackage[colorlinks=true,linkcolor=blue,filecolor=blue,urlcolor=blue,citecolor=blue]{hyperref}
\usepackage{footnote}
\usepackage[T1]{fontenc}

\begin{document}


\title{Level structure of light neutron-rich La isotopes beyond the N=82 shell closure}


\author{A.~Navin}
\thanks{Corresponding author}
\affiliation{GANIL, CEA/DRF-CNRS/IN2P3, Bd Henri Becquerel, BP 55027, F-14076 Caen Cedex 5, France}

\author{E.~H.~Wang}
\thanks{Corresponding author}
\affiliation{Shandong Provincial Key Laboratory of Nuclear Science, Nuclear Energy Technology and Computer Utilization, Weihai Frontier Innovation Institute of Nuclear Technology, School of Nuclear Science, Energy and Power Engineering, Shandong University, Jinan 250061, China.}
\affiliation{Weihai Research Institute of Industrial Technology of Shandong University, Weihai 264209, China}
\affiliation{Department of Physics and Astronomy, Vanderbilt University, Nashville, Tennessee 37235, USA}

\author{S.~Bhattacharyya}
\affiliation{Variable Energy Cyclotron Centre, 1/AF Bidhannagar, Kolkata 700064, India}
\affiliation{Homi Bhabha National Institute, Training School Complex, Anushaktinagar, Mumbai-400094, India}

\author{Menglan Liu}%
\affiliation{Sino-French Institute of Nuclear Engineering and Technology, Sun Yat-sen University, Zhuhai 519082, China}

\author{Cenxi Yuan}%
\affiliation{Sino-French Institute of Nuclear Engineering and Technology, Sun Yat-sen University, Zhuhai 519082, China}

\author{M.~Rejmund}
\affiliation{GANIL, CEA/DRF-CNRS/IN2P3, Bd Henri Becquerel, BP 55027, F-14076 Caen Cedex 5, France}

\author{A.~Lemasson}
\affiliation{GANIL, CEA/DRF-CNRS/IN2P3, Bd Henri Becquerel, BP 55027, F-14076 Caen Cedex 5, France}

\author{S.~Biswas}
\altaffiliation{Present address: Rutherford Appleton Laboratory, Fermi Ave, Harwell,
  Didcot OX11 0QX, United Kingdom}
\affiliation{GANIL, CEA/DRF-CNRS/IN2P3, Bd Henri Becquerel, BP 55027, F-14076 Caen Cedex 5, France}

\author{Y.H.~Kim}
\altaffiliation{Present address: Center for Exotic Nuclear Studies, Institute for Basic Science, Daejeon 34126, Republic of Korea}
\affiliation{Center for Exotic Nuclear Studies, Institute for Basic Science, Daejeon 34126, Republic of Korea}

\author{C.~Michelagnoli}
\altaffiliation{Present address: Institut Laue-Langevin, F-38042 Grenoble Cedex, France}
\affiliation{GANIL, CEA/DRF-CNRS/IN2P3, Bd Henri Becquerel, BP 55027, F-14076 Caen Cedex 5, France}

\author{J.~H.~Hamilton}
\affiliation{Department of Physics and Astronomy, Vanderbilt University, Nashville, Tennessee 37235, USA}

\author{A.~V.~Ramayya}
\affiliation{Department of Physics and Astronomy, Vanderbilt University, Nashville, Tennessee 37235, USA}

\author{I.~Stefan}
\affiliation{Université Paris-Saclay, CNRS/IN2P3, IJCLab, 91405 Orsay, France}

\author{R.~Banik}
\altaffiliation{Present address: Institute of Engineering and Management, Saltlake Sector V, Kolkata 700091, India}
\affiliation{Variable Energy Cyclotron Centre, 1/AF Bidhannagar, Kolkata 700064, India}

\author{P.~Bednarczyk}
\affiliation{Institute of Nuclear Physics PAN, 31-342 Krak\'ow, Poland}

\author{Soumik~Bhattacharya}
\affiliation{Variable Energy Cyclotron Centre, 1/AF Bidhannagar, Kolkata 700064, India}

\author{E.~Cl\'{e}ment}
\affiliation{GANIL, CEA/DRF-CNRS/IN2P3, Bd Henri Becquerel, BP 55027, F-14076 Caen Cedex 5, France}

\author{H.L.~Crawford}
\affiliation{Nuclear Science Division, Lawrence Berkeley National Laboratory, Berkeley, California 94720, USA}

\author{G.~de~France}
\affiliation{GANIL, CEA/DRF-CNRS/IN2P3, Bd Henri Becquerel, BP 55027, F-14076 Caen Cedex 5, France}

\author{P.~Fallon}
\affiliation{Nuclear Science Division, Lawrence Berkeley National Laboratory, Berkeley, California 94720, USA}

\author{G.~Fr\'{e}mont}
\affiliation{GANIL, CEA/DRF-CNRS/IN2P3, Bd Henri Becquerel, BP 55027, F-14076 Caen Cedex 5, France}

\author{J.~Goupil}
\affiliation{GANIL, CEA/DRF-CNRS/IN2P3, Bd Henri Becquerel, BP 55027, F-14076 Caen Cedex 5, France}

\author{B.~Jacquot}
\affiliation{GANIL, CEA/DRF-CNRS/IN2P3, Bd Henri Becquerel, BP 55027, F-14076 Caen Cedex 5, France}

\author{H.J.~Li}
\affiliation{GANIL, CEA/DRF-CNRS/IN2P3, Bd Henri Becquerel, BP 55027, F-14076 Caen Cedex 5, France}

\author{J.~Ljungvall}
\affiliation{Université de Strasbourg, CNRS, IPHC UMR 7178, 67000, Strasbourg, France}
\affiliation{Université Paris-Saclay, CNRS/IN2P3, IJCLab, 91405 Orsay, France}

\author{Y.~X.~Luo}
\affiliation{Department of Physics and Astronomy, Vanderbilt University, Nashville, Tennessee 37235, USA}
\affiliation{Lawrence Berkeley National Laboratory, Berkeley, California 94720, USA}

\author{A.~Maj}
\affiliation{Institute of Nuclear Physics PAN, 31-342 Krak\'ow, Poland}

\author{L.~M\'enager}
\affiliation{GANIL, CEA/DRF-CNRS/IN2P3, Bd Henri Becquerel, BP 55027, F-14076 Caen Cedex 5, France}

\author{V.~Morel}
\affiliation{GANIL, CEA/DRF-CNRS/IN2P3, Bd Henri Becquerel, BP 55027, F-14076 Caen Cedex 5, France}

\author{G.~Mukherjee}
\affiliation{Variable Energy Cyclotron Centre, 1/AF Bidhannagar, Kolkata 700064, India}
\affiliation{Homi Bhabha National Institute, Training School Complex, Anushaktinagar, Mumbai-400094, India}

\author{R.~Palit}
\affiliation{Department of Nuclear and Atomic Physics, Tata Institute of Fundamental Research,Mumbai, 400005, India}

\author{R.M.~P\'erez-Vidal}
\affiliation{Instituto de F\'isica Corpuscular, CSIC-Universitat de Val\`encia, E-46980 Val\`encia, Spain}

\author{J.~O.~Rasmussen}
\affiliation{Lawrence Berkeley National Laboratory, Berkeley, California 94720, USA}

\author{J.~Ropert}
\affiliation{GANIL, CEA/DRF-CNRS/IN2P3, Bd Henri Becquerel, BP 55027, F-14076 Caen Cedex 5, France}

\author{C.~Schmitt}
\altaffiliation{Present address: IPHC Strasbourg, Universit\'e de Strasbourg-CNRS/IN2P3, F-67037 Strasbourg Cedex 2, France}
\affiliation{GANIL, CEA/DRF-CNRS/IN2P3, Bd Henri Becquerel, BP 55027, F-14076 Caen Cedex 5, France}

\author{S.~J.~Zhu}
\affiliation{Department of Physics, Tsinghua University, Beijing 100084, People's Republic of China}


\date{\today}

\begin{abstract}

  The high spin excited states of Lanthanum isotopes $^{140-143}$La,
  above the $N=82$ closed shell, have been populated in fission reactions. The
  prompt $\gamma$-ray transitions were  measured using two complementary methods;
  a) in coincidence with the isotopically identified fragments produced in the fission
  of the $^{238}$U+$^{9}$Be system using the VAMOS++ and the AGATA spectrometers and
  b) high statistics three-fold $\gamma-\gamma-\gamma$ and four-fold $\gamma-\gamma-\gamma-\gamma$ coincidence
  data from the spontaneous fission of $^{252}$Cf using the Gammasphere.
  This work reports the first identification of a pair of parity doublet structures
  in $^{143}$La and the new high spin level structure in $^{140-142}$La from prompt $\gamma$-ray
  spectroscopy.
  The level structures are interpreted in terms of the systematics of neighbouring odd-$Z$
  nuclei above $Z=50$ shell closure and large-scale shell model calculations.
  The present results indicate the  presence of stable octupole deformation, in $^{143}$La.
  The excitation energy pattern and their comparison with neighbouring isotones,
  moving away from the N=82 closed shell, point towards a
  transition from single particle structures to an alternating parity
  rotational band structure in the La isotopic chain.

\end{abstract}


\maketitle


\section{Introduction}

Nuclei with few neutrons and protons above major shell closures are always of
special interest as they provide important inputs regarding single particle configurations,
neutron-proton  interactions, shape evolution, etc., and also to improve the
effective interaction for large scale shell model calculations.
In particular, the odd-A or odd-odd nuclei with respect to the even-even core
are important because of the important role played by the odd valence particles 
in determining their level structure.
As more and more valence protons are added to the $Z=50$ core, the nuclei begin
to show a transition from near spherical to a deformed structure.
Deformed nuclei can have both axial and reflection symmetry
(prolate or oblate deformation), or can show reflection asymmetric
shapes (manifested as octupole deformation)~\cite{Butler1996, Ahmad1993, Gaffney2013}.
The existence  of an "island of octupole collectivity" around Z = 56 and N =88
is confirmed by extensive experimental observation of parity doublets  in
various isotopes  in this region~\cite{Zhu2012,Li2012,Wang2015,Urb2016,Huang2016,Zhu2020,Urb2012,Huang2017,Luo2010,Wang2017,Bre2023,Wang2021,Rza2012}.
The nuclei in this region are suitable candidates to show octupole
collectivity, as their Fermi levels lie between the
$f_{7/2}$ and $i_{13/2}$ neutron orbitals and $d_{5/2}$ and $h_{11/2}$
proton orbitals, which differ by $j=l=3$~\cite{Leander1985,Naz85}.
Various experimental signatures to characterize stable octupole deformation 
include features of rotational bands~\cite{Butler1996, Afana1995, Ahmad1993}.
The existence of parity doublet bands (pair of the states with spin I but
of opposite parities and having strong interband E1 transitions) is one such
signature.
These bands are associated with the simplex quantum numbers $s=\pm 1$ for even-even
nuclei, while for odd-A nuclei it is $s= \pm i$.
However, experimental observation of two sets of parity doublet bands has been
reported only in a few even-even ~\cite{Huang2016, Chen2006} and
odd-A~\cite{Zhu1999, Chakraborty2023} nuclei. Particularly, for odd-Z nuclei, it is not yet reported.


The La ($Z=57$) isotopes with $A\sim140$, the seven valence protons
above $Z=50$ closed shell show interesting features of the evolution from prolate
deformation to the onset of octupole collectivity, as a function of
neutron number.
The neutron-rich La ($Z=57$) isotopes are predicted 
to lie within a region of mixed quadrupole-octupole
deformation~\cite{nazare1984, Naz85,Pie85}.
A competition between the symmetric and asymmetric shapes has also been pointed
out in $^{145}$La~\cite{Zhu1999}.
Above $N=82$, $^{140-142}$La isotopes
are expected to lie in a transitional region, where the interplay of prolate
deformation and octupole collectivity
is expected  to evolve. On the other hand, $^{143}$La (N=86) lies near the lower edge
of the  neutron number of this island, where the presence of octupole
collectivity was reported~\cite{Luo2009}.
The octupole collectivity is expected to  increase towards $N= 88-90$ (region of
maximum collectivity), however in $^{146,147}$La ($N=89,90$), it is  
reported to be weakened~\cite{Urb1996}. 
Thus, to understand the onset of octupole deformation in the La isotopic chain and
to search for parity doublet bands corresponding to both simplex quantum numbers,
investigations on high-spin states in the La isotopes above $N=82$ are necessary.
While the heavier neutron-rich La isotopes have been studied and characterized
through high-fold $\gamma$-ray  spectroscopy produced in $^{252}$Cf spontaneous fission,
the lighter La isotopes just above $N=82$ are difficult to access.
In particular, the high spin states in $^{140,141}$La from in-beam spectroscopy,
have not yet been characterized and
several ambiguities exist on the high spin structures of $^{142,143}$La.
Recent measurements  of neutron separation energies obtained from new
precision mass measurement \cite{jaries2024} have shown the strongest
change of two-neutron separation energy in the periodic table between $N=92$
and $N=94$ in La nuclei indicating more interesting aspects in this isotopic chain.

Here, we report the first identification of a pair of parity doublet structures in $^{143}$La
and the high-spin level structures of $^{140-142}$La isotopes. The excited high spin states of
these isotopes were measured using $\gamma$-ray spectroscopy, 
through two complementary techniques, combining isotopic identification
of the fission fragments and high-fold $\gamma$-ray correlations.
Further, the previously reported high-spin states in $^{142}$La~\cite{Luo2021}
are now reassigned to $^{143}$La based on unambiguous isotopic identification.

\section{Experiment}

The excited states and the $(A,Z)$ identified $\gamma$ rays of $^{140-143}$La isotopes have been
investigated using two complimentary fission experiments, {\it viz.},
(i) prompt $\gamma$-ray spectroscopy from fusion-fission and transfer-induced
fission reactions in inverse kinematics,
using a $^{238}$U beam of energy $6.2$~MeV/u, on a $^{9}$Be target (thicknesses 
$1.6$~$\mu$m and $5$~$\mu$m) at Grand Acc\'el\'erateur National d'Ions Lourds (GANIL), France;
and (ii) high statistics three-fold $\gamma-\gamma-\gamma$ and four-fold
$\gamma-\gamma-\gamma-\gamma$ coincidence data from the spontaneous fission
of $^{252}$Cf using the Gammasphere, at LBNL, USA.
These two complementary methods were combined earlier for the study of
neutron-rich Pr~\cite{Wang2015}, Pm~\cite{Bhattacharyya2018} and Y~\cite{Wang20212} isotopes. 

In the  experiment at GANIL, the $(A, Z)$ identification of the fission fragments was obtained 
using the  large acceptance magnetic spectrometer
VAMOS++~\cite{Rejmund2011, Alahari2014}.
It was placed at an angle of $20^{\circ}$ with respect to the beam
axis~\cite{Alahari2014, Rejmund2011, Kim2017}.
The $Z$ identification was obtained from $\Delta E$-$E$, using the ionization chamber at
the focal plane of VAMOS++. The time of flight of each fragment was measured between a Dual
Position-Sensitive Multi-Wire Proportional Counter (DPS-MWPC)~\cite{Vandebrouck2016},
placed behind the target at the entrance of VAMOS++, and the MWPC placed at the focal plane. 
The focal plane detector system also includes two drift chambers (DC)
for tracking the trajectory of the ions.
The mass number ($A$), atomic number ($Z$), atomic charge ($Q$) and the velocity 
of the fission fragments were determined on an event-by-event basis.
The mass-to-charge ratio $(M/Q)$ was obtained from the reconstructed magnetic rigidity
(B$\rho$) and the velocity of the fragment.  
The prompt $\gamma$ rays, emitted in coincidence with a given fission fragment,
were detected using the AGATA setup at GANIL~\cite{Clement2017},
placed 13.5~cm from the target. The Doppler correction for the $\gamma$ rays was made
using the velocity of the emitting fragments and the corresponding position of the $\gamma$ rays in AGATA. The $\gamma$-ray energy uncertainties after Doppler correction,
are typically $\pm$ 0.2~keV, $\pm$ 0.5~keV and $\pm$ 1~keV around 200~keV,
500~keV and 1~MeV respectively.
The data were collected demanding a coincidence  between the prompt $\gamma$ rays
and DPS-MWPC conditioned by the MWPC in the focal plane within $300$~ns.
Further details and identification of fragments and corresponding coincidence $\gamma$ rays
in AGATA are given in Ref.~\cite{Kim2017, Banik2020}.

The high-fold $\gamma$-coincidence data was obtained using 101 HPGe  detectors of
the Gammasphere setup at LBNL. A 62-$\mu$Ci $^{252}$Cf source
was sandwiched between two Fe foils of thickness $10$~mg/cm$^2$.
The fission fragments and $\alpha $ particles were stopped by foils, while neutrons 
and $\beta$ particles  were partially moderated and absorbed by the foils and
a 7.6 cm diameter plastic ball. 
The data were sorted into $\gamma$-$\gamma$-$\gamma$  and higher fold $\gamma$
events to form a cube (three-fold) and more recently in a  hyper-cube (four-fold), and were  analyzed 
using the RADWARE package~\cite{Radford1995}.
More details of the experiment and analysis procedures can be found in~\cite{Wang2015}.

\section{Results}

\begin{figure}[t!]
 \includegraphics[width=\columnwidth,clip=true]{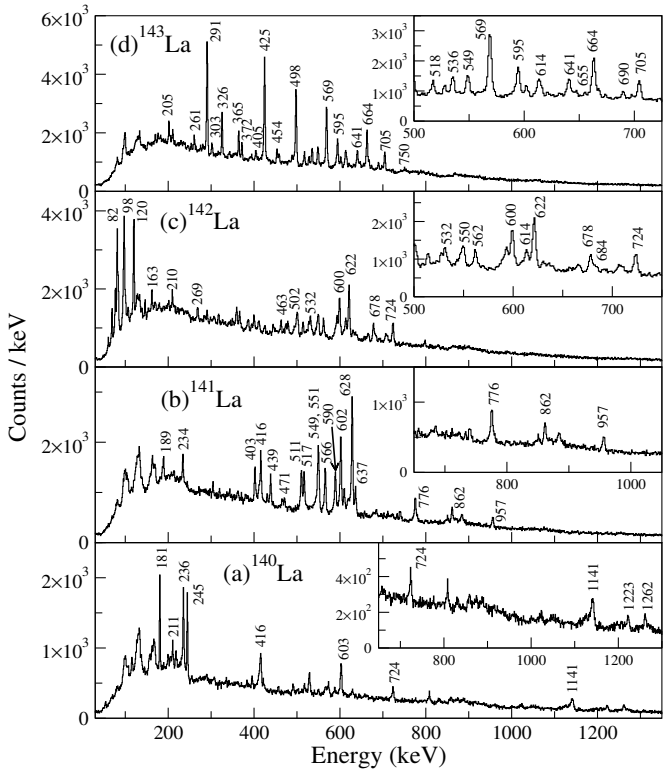}
 \caption{(A,Z) gated Doppler corrected ``singles'' $\gamma$-ray spectra of
   $^{140-143}$La, detected in AGATA, from the fragment-$\gamma$ coincidences in
   $^{238}$U+$^{9}$Be-induced fission data. The insets show part of the spectra
   in an expanded scale. }
 \label{Singles}
\end{figure}

The present work reports on the in-beam prompt $\gamma$ rays from high spin states
of isotopically identified $^{140-143}$La nuclei. 
The $\gamma$ rays decaying from the low-lying states of $^{140-142}$La have been known
earlier either from $\beta$-decay~\cite{Meyer1990, Rufino2022, Chung1983}
or from $(n, \gamma)$ reactions~\cite{Jurney1970}.
For $^{143}$La, the high spin states are known from
the study of spontaneous fission of $^{252}$Cf~\cite{Wang2007, Luo2009}. 
The isotopically identified Doppler corrected $\gamma$-ray spectra
of $^{140-143}$La isotopes, 
observed in the present work using AGATA, are shown in \figurename~\ref{Singles}.
In the case of $^{142-143}$La, it was additionally
possible to use the high fold $\gamma$ coincidence cube and
hypercube data, obtained using Gammasphere. 
The details of the  level structure of the La isotopes reported
in the present work are given in the following subsections.

\subsection{$^{140}$La}

The present work reports on the first measurements of $\gamma$ rays from
high-spin states of $^{140}$La.
The previous studies on $^{140}$La reported the $\gamma$ rays depopulating mostly
the low spin excited states, observed either in $\beta$-decay~\cite{Meyer1990}
or from (n,$\gamma$) reactions~\cite{Jurney1970}. 

\begin{figure}[ht!]
       \includegraphics[width=0.8\columnwidth,angle=-90,clip=true]{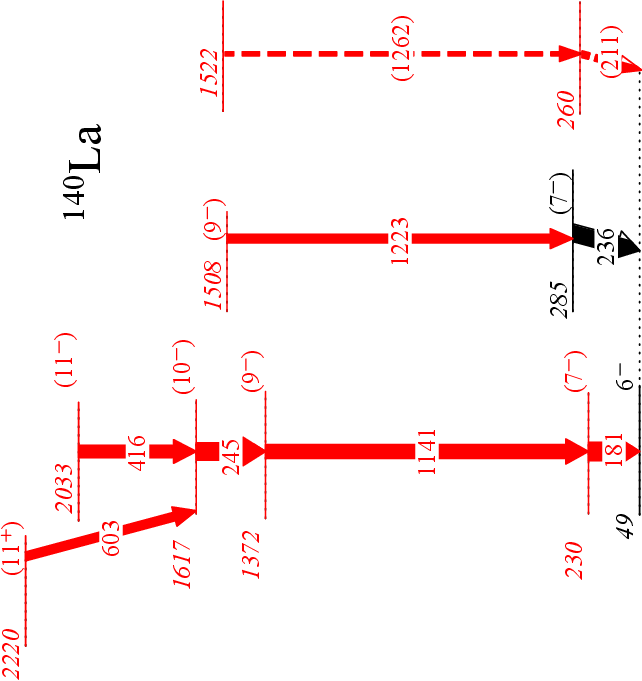}
 \caption{Level scheme of $^{140}$La, obtained from the present work.
   The $6^-$
   level was reported in $\beta$-decay~\cite{Meyer1990} and (n,$\gamma$) reactions~\cite{Jurney1970}.
   The new transitions from the present work are marked in red colour. The 211 and $1262$~keV $\gamma$ transitions are denoted as dashed lines to indicate the  tentative nature of their placement (see text). Their assignment to $^{140}$La is confirmed from the (Z,A) gated spectrum.
 }
 \label{LS_140La}
\end{figure}

The Doppler corrected $\gamma$-ray spectrum of $^{140}$La, measured in AGATA,
is shown in \figurename~\ref{Singles}(a).
The proposed level scheme of $^{140}$La, as obtained from the present work,
is shown in \figurename~\ref{LS_140La}. The energies shown are adopting the placement of $235.8$~keV
transition in Ref.~\cite{Jurney1970}, and following the energy systematics
of neighbouring $N=83$ isotone $^{138}$Cs.
The $\gamma$ rays, assigned to $^{140}$La, along with their relative intensities 
and the level energies, obtained from the (A,Z) gated spectra are shown in 
Table.~\ref{tab:Table1} in the appendix. 

\begin{figure}[t!]
  \includegraphics[width=\columnwidth,clip=true]{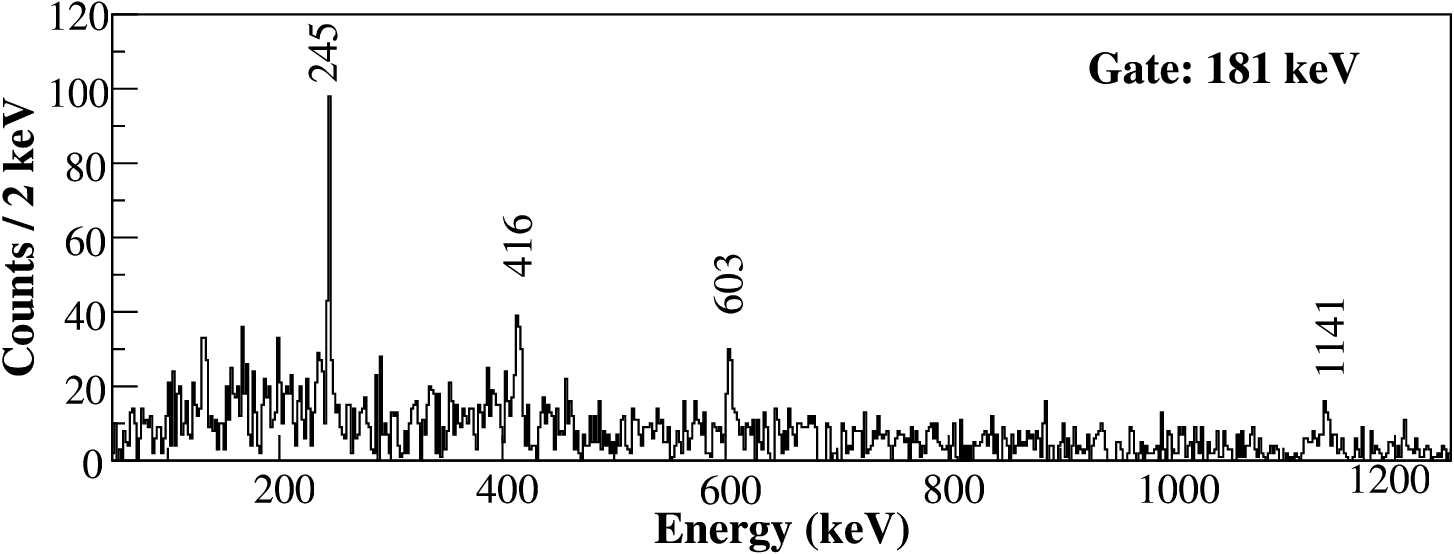}
  \includegraphics[width=\columnwidth,clip=true]{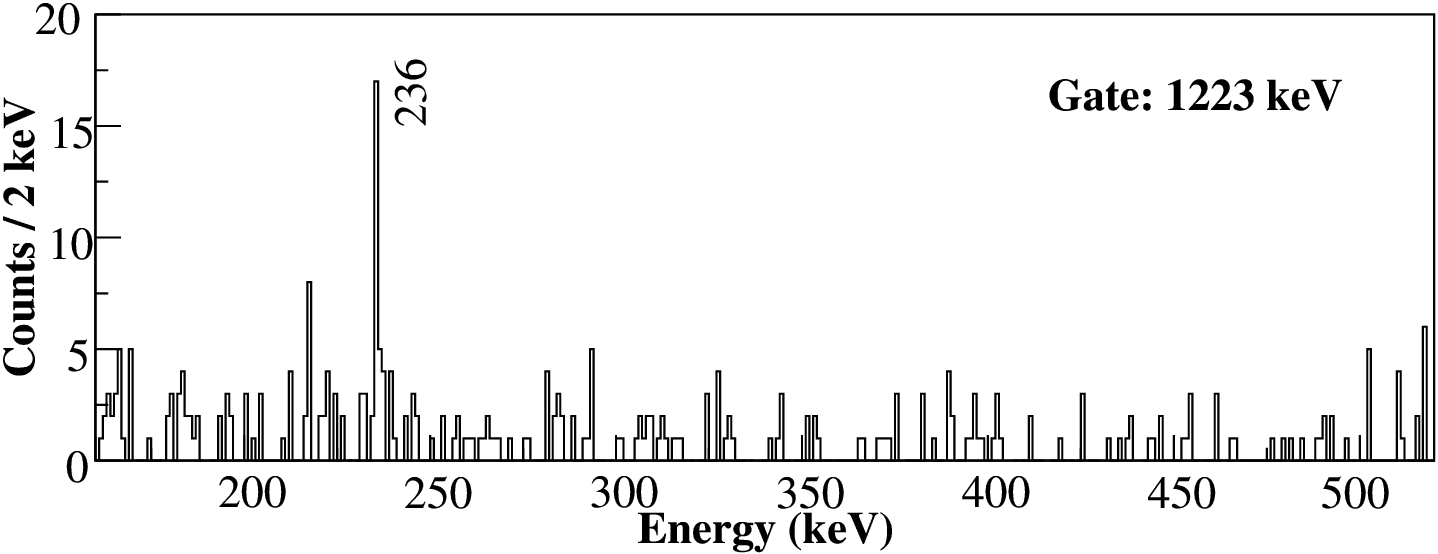}
  \caption{Coincidence spectrum corresponding to the $181$~keV (upper panel) and
    $1223$~keV (lower panel) $\gamma$-rays of $^{140}$La,
    obtained from the $^{238}$U+$^{9}$Be-induced fission data. }
 \label{140La-gate}
\end{figure}

The prompt coincidence spectra with coincidence gates on the $181$~keV
and $1223$~keV $\gamma$ rays, obtained using AGATA, are shown in \figurename~\ref{140La-gate}.
The presence of the strong $245$, $416$, $603$ and $1141$~keV transitions
can be clearly seen. The $603$~keV transition is found to be in coincidence
with all these transitions, except the $416$~keV transition.
Thus, the $181$, $1141$, $245$ and $416$~keV
transitions are placed in a cascade, on the basis of their mutual coincidence
relations and relative intensities, whereas, the 603~keV transition is placed in
parallel with the 416~keV transition.
The intense $236$~keV $\gamma$-ray, observed in the singles spectrum, is found
to be in coincidence only with the $1223$~keV $\gamma$-ray, as shown in
\figurename~\ref{140La-gate}. Therefore, the $236-1223$~keV coincidence cascade is
placed in parallel with the $181-1141$~keV band $\gamma$ rays.
The other two weak transitions $1262$ and $211$~keV,
were found to have a weak coincidence only with each other and
therefore are tentatively placed as a parallel cascade. Due to this reason the 211 and $1262$~keV transitions are denoted as dashed lines. Though their assignment to $^{140}$La is confirmed from the mass gated spectrum, but their placement in level scheme is tentative due to lack of coincidence statistics.
The $724$~keV $\gamma$ ray, observed in the singles $\gamma$-ray spectrum
(\figurename~\ref{Singles}(a)), obtained in coincidence with the $^{140}$La fragment,
does not show any coincidence with the other observed $\gamma$ rays.
A $722.5$~keV transition was reported in a (n,$\gamma$) study~\cite{Jurney1970},
as decaying to the $48.8$~keV $6^-$ level. The $724$~keV $\gamma$ ray, observed in
the present work, could be the same transition, decaying directly to the $6^-$
level in parallel with the other transitions. 


Earlier studies on $^{140}$La from thermal neutron capture
reactions~\cite{Jurney1970} also reported two parallel transitions
of $180.8$ and $235.8$~keV, as decaying from the $284.6$~keV level,
to the two different $6^-$ levels at $103.8$~keV and $48.8$~keV, respectively.
However, these two $\gamma$ rays were not observed in $\beta$-decay~\cite{Meyer1990}.
The study of low-lying states of $^{140}$La by $(d,p)$
reaction~\cite{Kern1966} also reported the states at $284.6$~keV
and $48.8$~keV and assigned the spin-parity of $7^-$ and $6^-$, respectively. 
It is found that, the intensity of $180.8$~keV $\gamma$ ray with respect to
that of $235.8$~keV $\gamma$ ray is only 5\%, as reported
in Ref.~\cite{Jurney1970}. This intensity ratio is much less compared to
that of the $181$~keV $\gamma$-ray, observed in the present work (produced with almost equal
intensity as that of the $235.8$~keV transition).
Thus, the $181$~keV transition, observed in the present work and the $180.8$~keV,
observed in Ref.~\cite{Jurney1970}, are most likely not the same.
Further, from the present $\gamma-\gamma$ coincidence data,
the $181$ and $236$~keV transitions cannot be placed as decaying from the same level,
as the $\gamma$ rays of
$245$, $416$, $603$ and $1141$~keV, placed above the $181$~keV $\gamma$-ray,
are found to be in coincidence only with $181$~keV, but not with $236$~keV $\gamma$-ray. 
Therefore, in the present work, the $181$ and $236$~keV $\gamma$ rays
are placed as decaying from two different levels.
The $181$~keV transition, observed in the present work, is placed as decaying to
the same $6^-$ level at $48.8$~keV, though its placement as decaying to
any of the other known low-lying levels in $^{140}$La cannot be ruled out. 
The $236$~keV $\gamma$-ray is considered to decay from a $7^-$ level
to a $6^-$ level, as reported in Ref.~\cite{Jurney1970, Kern1966}.
The spin-parity assignments of the state decaying by $181$~keV transition
and the higher spin states corresponding to the $\gamma$-ray cascades
of $245$, $416$, $603$ and $1141$~keV, are made on the basis of the systematics
of the neighbouring $N=83$ isotones $^{138}$Cs~\cite{Liu2010}. In the present work, the induced fission data can only determine the energy levels and intensities of the transitions. The calculations discussed later in the text support the tentative spin/parity assignments that have been made.

\subsection{$^{141}$La}

The present work reports on the in-beam $\gamma$ rays from high spin states in
$^{141}$La, for the first time. Prior to the present measurements, 
the information on the excited states of $^{141}$La at low spin was
mainly obtained from the $\beta$-decay of $^{141}$Ba~\cite{Faller1986, Rufino2022}.
A recent work~\cite{Rufino2022} reports
detailed measurements of $\gamma-\gamma$ coincidence, angular correlation
of decay $\gamma$ rays and $\beta$-feeding to low-lying positive and
negative parity states. None of these $\gamma$ rays of $^{141}$La
are observed in the present study.

\begin{figure}[ht!]
    \includegraphics[height=\columnwidth,angle=-90,scale=0.75,clip=true]{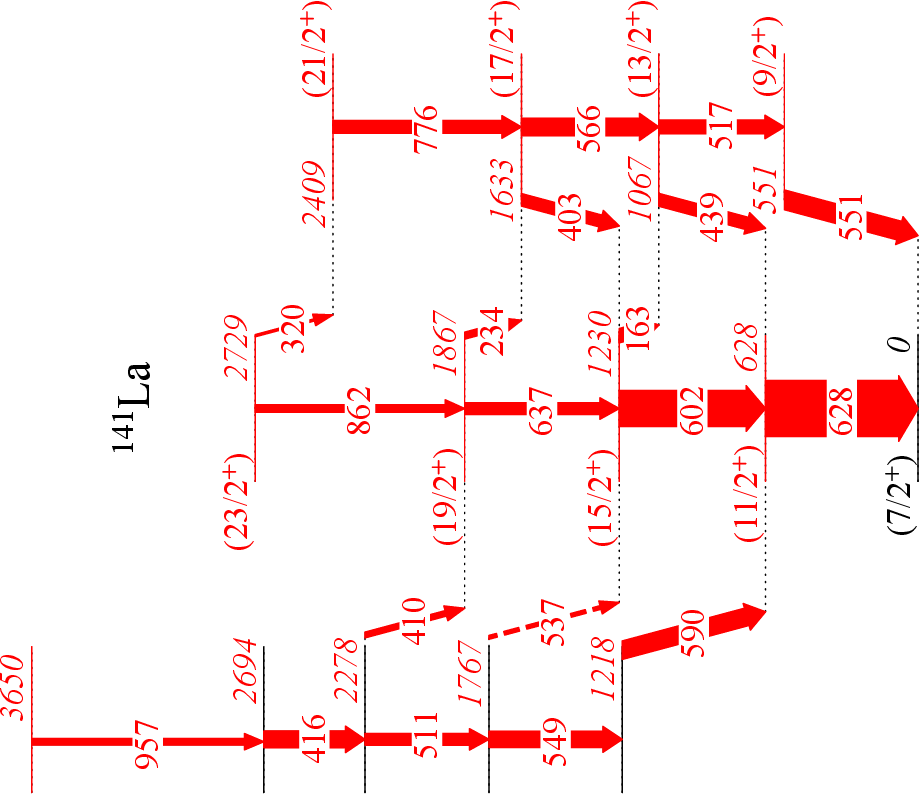}
    \caption{Level scheme of $^{141}$La, obtained from the present work.
    The new transitions from the present work are marked in red colour.}
 \label{LS_141La}
\end{figure}

\figurename~\ref{LS_141La} shows the proposed level scheme of $^{141}$La, obtained
from the present work. The $\gamma$ rays placed in the level scheme are 
identified on the basis of the $\gamma$ rays detected in coincidence with the
$^{141}$La fragments. The corresponding $\gamma$ spectrum is shown in
\figurename~\ref{Singles}(b). A majority of the $\gamma$ rays observed in this
spectrum are placed in the level scheme (\figurename~\ref{LS_141La}), on the basis
of prompt $\gamma-\gamma$ coincidence and their relative intensities.
The $\gamma$ rays, assigned to $^{141}$La, along with their relative intensities 
and the level energies, obtained from the (A,Z) gated spectra are shown in 
Table.~\ref{tab:Table2} in the appendix. 

\begin{figure}[ht!]
 \includegraphics[width=\columnwidth,clip=true]{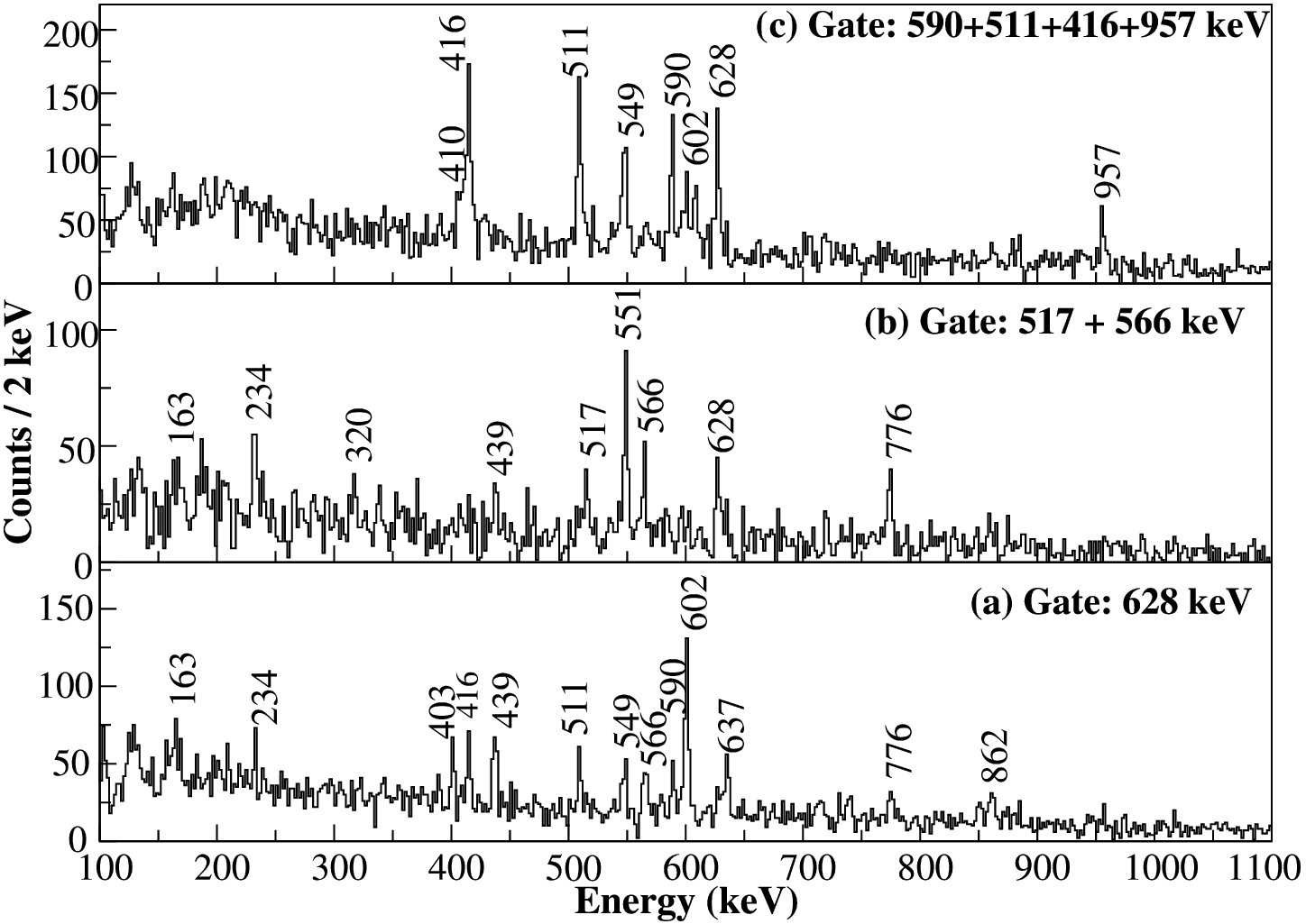}
 \caption{Coincidence spectrum corresponding (a) the 628~keV gate, (b) the added gates of
   517 and 566~keV $\gamma$ rays, (c) the added gates of 590, 511, 416 and 957~keV
   $\gamma$ rays of $^{141}$La, obtained from $^{238}$U+$^{9}$Be-induced fission data. }
 \label{141La-gate}
\end{figure}

The coincidence spectra corresponding to different $\gamma$-ray
coincidence conditions are shown in \figurename~\ref{141La-gate}(a-c).
The coincidence spectrum corresponding to the gate of the most intense
$\gamma$-ray of $628$~keV, observed in $^{141}$La, is shown in \figurename~\ref{141La-gate}(a).
It can be seen that except for the $551$ and $517$~keV transitions, which are placed in
parallel to the band built on the $628$~keV $\gamma$-ray, all other $\gamma$ rays observed in the
single spectrum are found to be in coincidence with the $628$~keV $\gamma$ ray.
The $628-602-637-862$~keV $\gamma$ rays form the main cascade in the level scheme
of $^{141}$La. 
The $549$ and $551$~keV transitions cannot be separated
in the singles spectrum (\figurename~\ref{Singles}(b)). But the
summed coincidence gates of $517$ and $566$~keV (\figurename~\ref{141La-gate}(b))
and the summed coincidence gates of $590-511-416-957$~keV $\gamma$-ray cascade
(\figurename~\ref{141La-gate}(c)), clearly separate $551$~keV and $549$~keV
transitions. Thus, the $549$~keV
and $551$~keV~ $\gamma$ rays are placed in two different cascades.
The set of transitions, consisting of $590-549-511-416-957$~keV show mutual coincidences,
and all these transitions are in coincidence with the $628$~keV $\gamma$-ray.
The $511$, $416$, and $957$~keV transitions also show a weak coincidence
with the $602$~keV $\gamma$-ray, which indicates a possible connection between the
two cascades. It is found that the $2278$~keV level is connected with
the $1867$~keV ($19/2^+$) state of the main yrast cascade through
a $410$~keV $\gamma$-ray.
However, the connection of the $1767$~keV state to the $1230$~keV ($15/2^+$) state
of the yrast cascade could not be established and thus is shown as a tentative transition
of $537$~keV in the level scheme (\figurename~\ref{LS_141La}). It may be noted that,
a similar cascade of transitions, connecting to each level of the yrast sequence,
has also been identified in the neighbouring $N=84$ isotone $^{139}$Cs~\cite{Liu2009}. 
The tentative spin-parity assignments of the new levels of the yrast band structures
in $^{141}$La are made on the basis of the systematics of $N=84$ isotones of Cs
and I nuclei~\cite{Liu2009}.

\subsection{$^{142}$La}

The present work reports on the de-excitation $\gamma$ rays from higher
spin states in $^{142}$La. The isotopically tagged prompt $\gamma$ rays in $^{142}$La
are shown in \figurename~\ref{Singles}(c).
The low spin states of $^{142}$La up to $1.5$~MeV excitation have been reported earlier following
the $\beta$- decay of $^{142}$Ba~\cite{Chung1983}. 
A set of cascade $\gamma$ rays was reported in Ref.~\cite{Luo2021}, as belonging to $^{142}$La,
from $^{252}$Cf spontaneous fission, but none of the $\gamma$ rays are observed in the
(A, Z) gated spectrum of the present work (\figurename~\ref{Singles}(c)).

\begin{figure}[ht]
    \includegraphics[height=\columnwidth,angle=-90,clip=true]{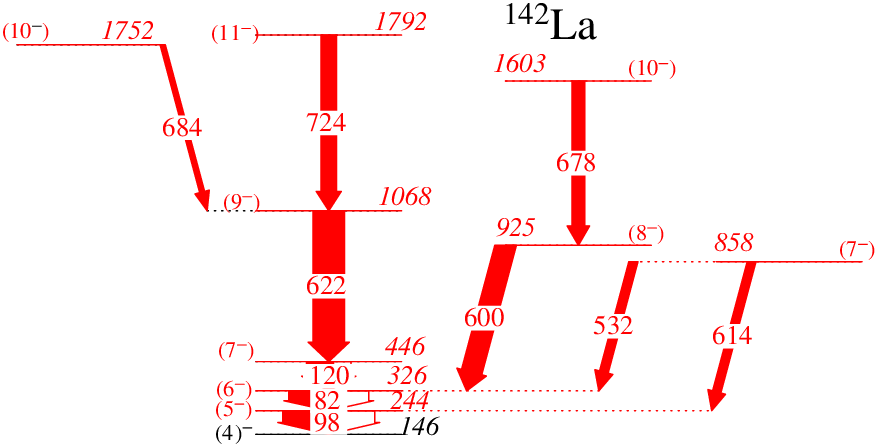}
 \caption{Level scheme of $^{142}$La, obtained from the present work.
   The $146$~keV ($4$)$^-$ level was
   reported in $\beta$-decay~\cite{Chung1983}.
   The new transitions from the present work are marked in red colour.
 }
 \label{LS_142La}
\end{figure}

The new level scheme of $^{142}$La, proposed from the present work is shown in 
\figurename~\ref{LS_142La}.
The $\gamma$ rays, assigned to $^{142}$La, along with their relative intensities 
and the level energies, obtained from the (A,Z) gated spectra are shown in 
Table.~\ref{tab:Table3} in the appendix. 

 

\begin{figure}[ht]
 \includegraphics[width=\columnwidth,clip=true]{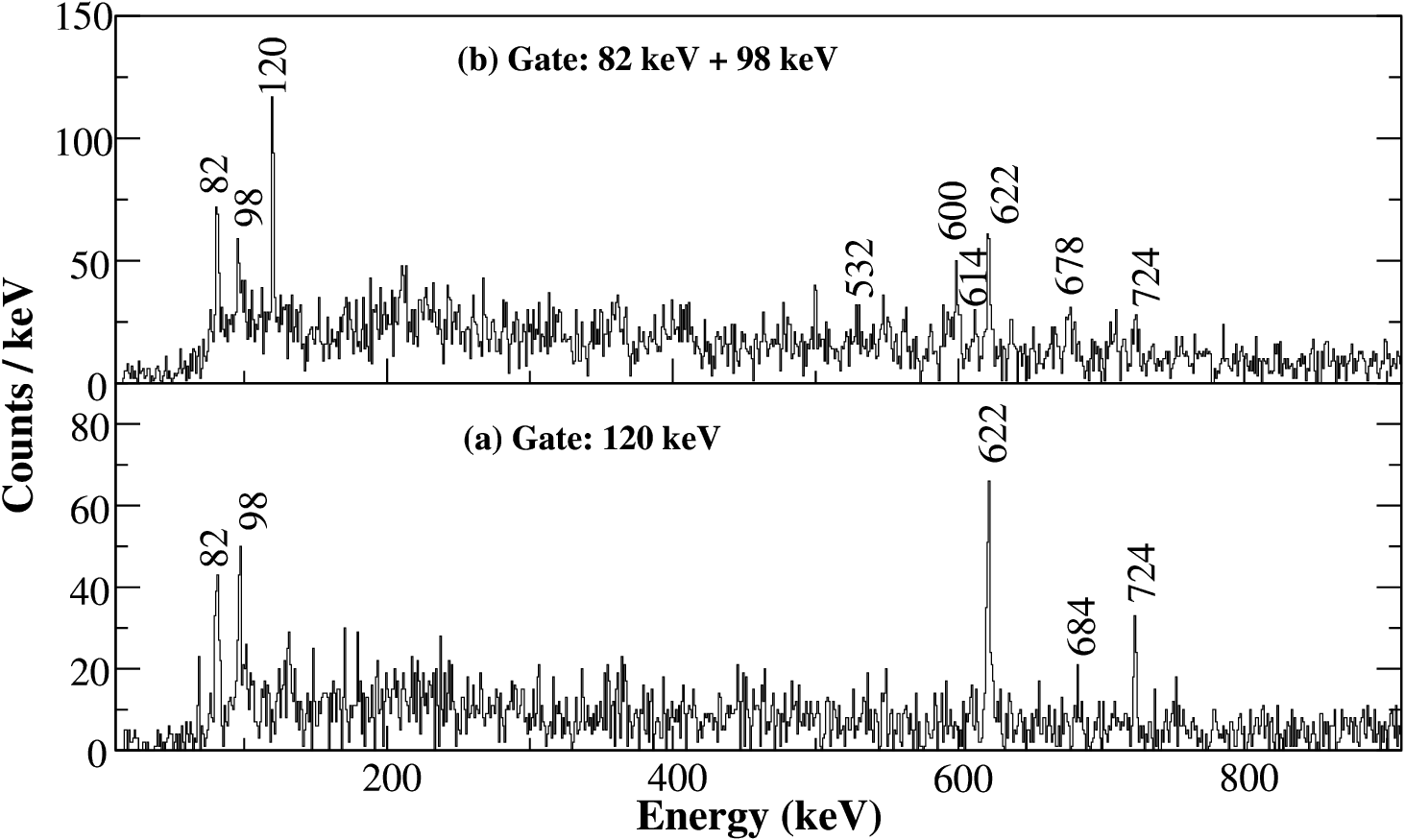}
 \caption{Coincidence spectra corresponding to the (a) added gates of 82 and 98 keV,
   (b) gate of 120 keV $\gamma$ rays in $^{142}$La, as obtained from the $^{238}$U+$^{9}$Be-induced fission data.}
 \label{142La-gate-agata}
\end{figure}

\begin{figure}[ht!]
 \includegraphics[width=\columnwidth,clip=true]{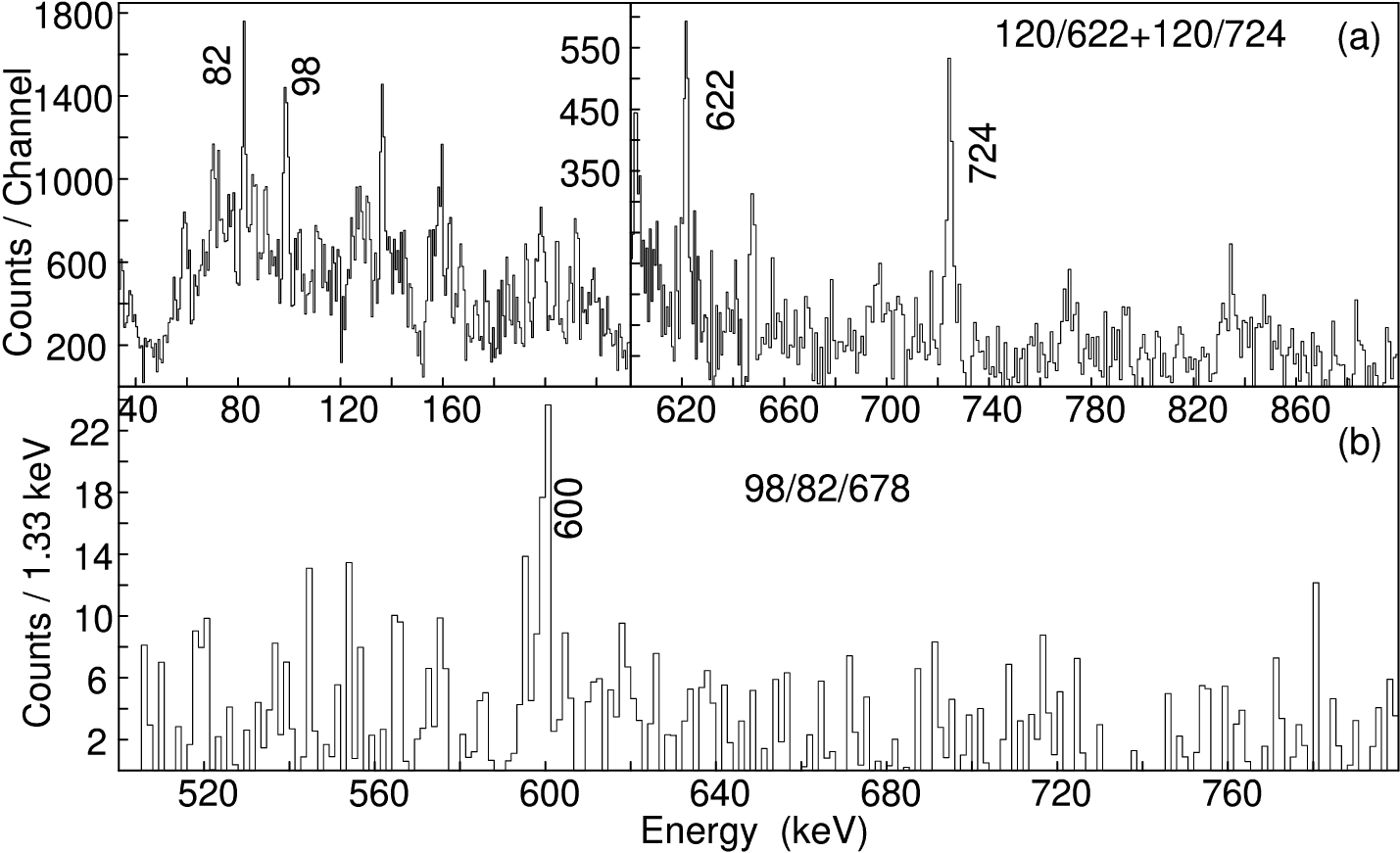}
 \caption{Coincidence spectra corresponding to the (a) added double gates of 120-622 and
   120-724, (b) triple gates of 98-82-678~keV $\gamma$ rays in $^{142}$La, 
 obtained from $^{252}$Cf fission using Gammasphere. }
 \label{142La-gate-Cf}
\end{figure}

\begin{figure*}[ht]
 \includegraphics[width=\textwidth,scale=0.8,clip=true]{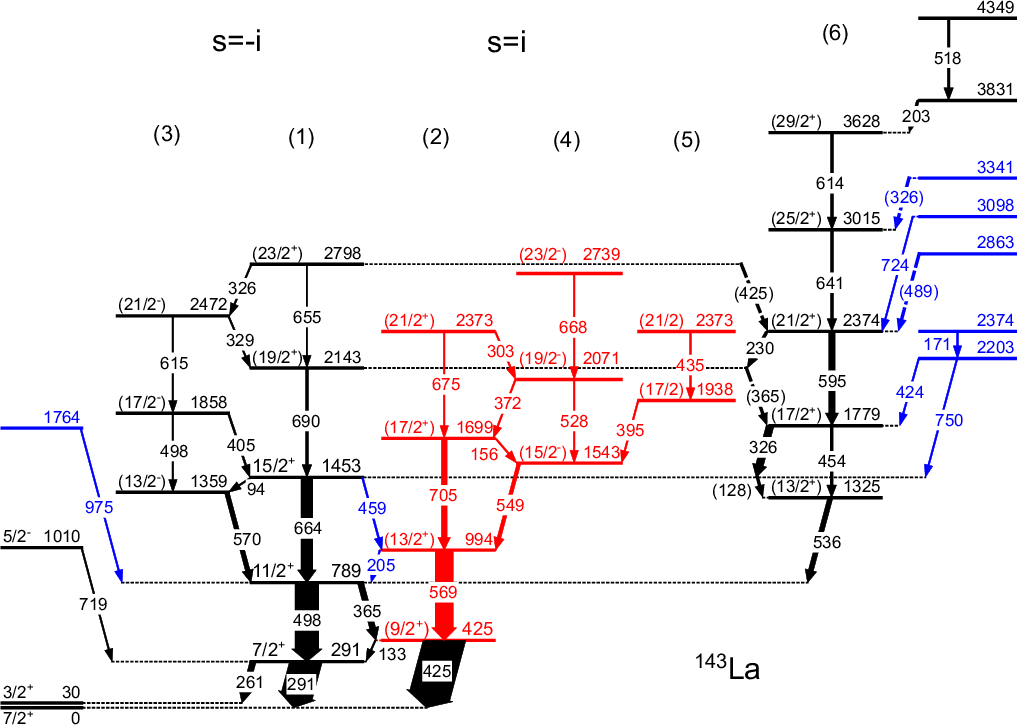}
 \caption{Level scheme of $^{143}$La, obtained from the present work.
   Transitions in black were reported in Ref.~\cite{Wang2007,Luo2009} from spontaneous
   fission of $^{252}$Cf.
   The $1010$~keV level and its decay by $719$~keV transition was reported only
   from $\beta$-decay~\cite{Faller1988}.
   Transitions in blue and red are newly assigned  $^{143}$La.
   The transitions in  red  were earlier misidentified as belonging to $^{142}$La in Ref.~\cite{Luo2021}.
 }
 \label{LS_143La}
\end{figure*}

A coincidence spectrum corresponding to the sum gate $82$ and $98$~keV $\gamma$ rays
using the data from AGATA is shown in \figurename~\ref{142La-gate-agata}(a).
It can be seen that most of the $\gamma$ rays of $^{142}$La are in coincidence with
$82$ and $98$~keV transitions. \figurename~\ref{142La-gate-agata}(b) shows the coincidence
spectrum of $120$~keV $\gamma$ ray, in which only the transitions corresponding to
the yrast cascade are present.
The coincidence relations of the other observed $\gamma$ rays, such as,
$502$, $550$ and $562$~keV, were not clear and therefore these are not placed in
the level scheme. However, their identification as belonging to the $^{142}$La is
confirmed by the fragment-$\gamma$ coincidence, shown in \figurename~\ref{Singles}(c).
In addition to the coincidence relationship  obtained using VAMOS-AGATA 
the high-fold $\gamma$-ray coincidences, obtained using Gammasphere
have  also been used. Such a combined analysis was only possible
for $^{142,143}$La isotopes in the present work due to the relatively lower yields of
$^{140,141}$La in the spontaneous fission of $^{252}$Cf. 
These high-fold $\gamma$-ray coincidence cube and hypercube,
made from the coincident data of the Gammasphere array have been utilized to further
confirm the placement of various cascade transitions in $^{142}$La.
\figurename~\ref{142La-gate-Cf}(a) shows the sumed double gates of $120/622$~keV
and $120/724$~keV transitions and,  the four-fold coincidence gates of
$98/82/678$~keV transitions from the hypercube data are shown in
\figurename~\ref{142La-gate-Cf}(b).
In part (a), it can be seen that the strong $82$, $98$, $622$ and
$724$~keV transitions, observed in the data from $^{238}$U+$^{9}$Be reaction
(\figurename~\ref{Singles} and \ref{142La-gate-agata}) are also confirmed by the $^{252}$Cf
spontaneous fission. The presence of the $600$~keV transition can be clearly seen
in the four-fold coincidence of the $98/82/678$ ~keV transitions.

In $^{252}$Cf spontaneous fission data, weak evidence of $\approx$ $68$ and $78$~keV
transitions are also seen. 
A previous $^{142}$Ba $\beta$-decay study~\cite{Chung1983}  reported the 68.3-77.6 and 69.7-77.6 keV cascades to the ground state depopulating the 145.9 and 147.3 keV levels,  respectively. In Ref.~\cite{Chung1983}, only the $146$~keV level was tentatively assigned to ($4$)$^-$. The level scheme shown in the Fig.~\ref{LS_142La} is built on this (4)$^-$ state.  
Based on the systematic comparison with $A=85$ isotones and
the shell model calculations discussed in the next section,
the tentative spin-parity of the excited states in $^{142}$La has been assigned. 


\subsection{$^{143}$La}

\subsubsection{Level Structure}

The level scheme of $^{143}$La  was earlier reported in Ref.~\cite{Wang2007,Luo2009}.
\figurename~\ref{LS_143La} shows the new and extended level scheme of $^{143}$La obtained in this work.
Bands (1), (3) and (6) in \figurename~\ref{LS_143La} were reported in Ref.~\cite{Wang2007,Luo2009}.
The strong transitions in these Bands are confirmed in the A and Z gated spectrum,
shown in \figurename~\ref{Singles}.
The $\gamma$ rays, assigned to $^{143}$La, along with their relative intensities 
and the level energies, obtained from the (A,Z) gated spectra are shown in 
Table.~\ref{tab:Table4} in the appendix.

\begin{figure}[t]
 \includegraphics[width=\columnwidth,scale=0.9,clip=true]{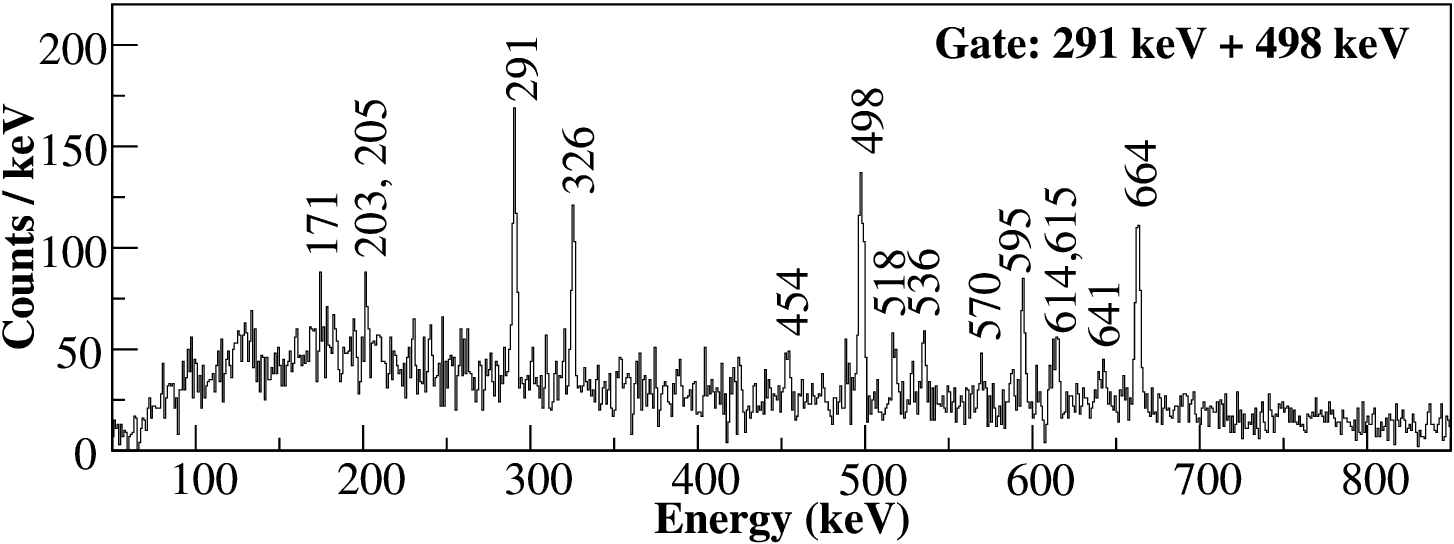}
 \caption{Coincidence spectra corresponding to the summed gates of the $291$
   and $498$~keV $\gamma$ rays 
 in $^{143}$La, as obtained from the $^{238}$U+$^{9}$Be induced-fission data.}
 \label{143La-gate-agata}
\end{figure}

\begin{figure}[t]
 \includegraphics[width=\columnwidth,scale=0.9,clip=true]{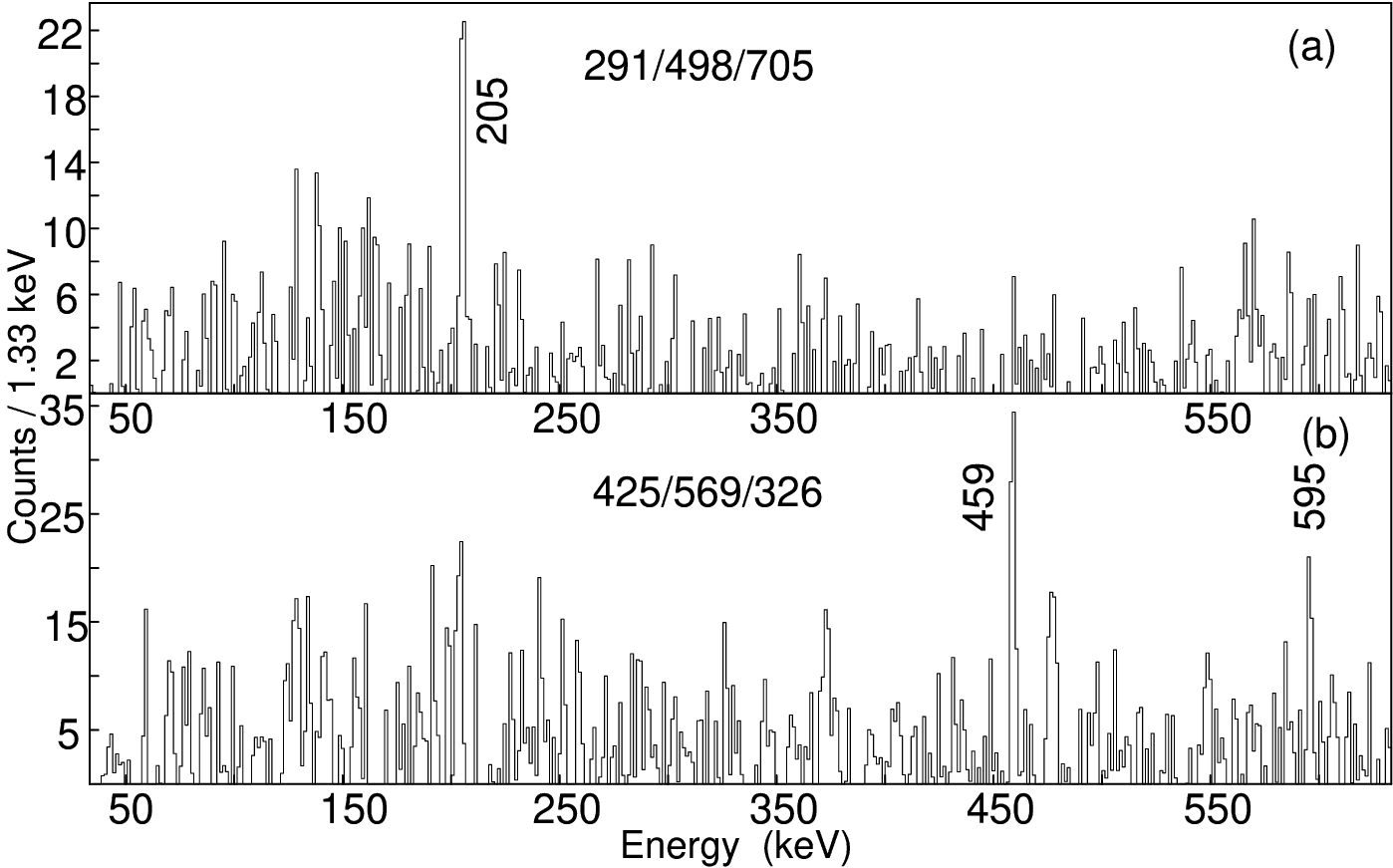}
 \caption{Coincidence spectra corresponding to the triple gates of (a) 291-498-705~keV
   and (b) 425-569-326~keV $\gamma$ rays in $^{143}$La, as obtained from the
   $^{252}$Cf fission data.}
 \label{143La-gate1-Cf}
\end{figure}

\begin{figure}[t]
 \includegraphics[width=\columnwidth,scale=0.95,clip=true]{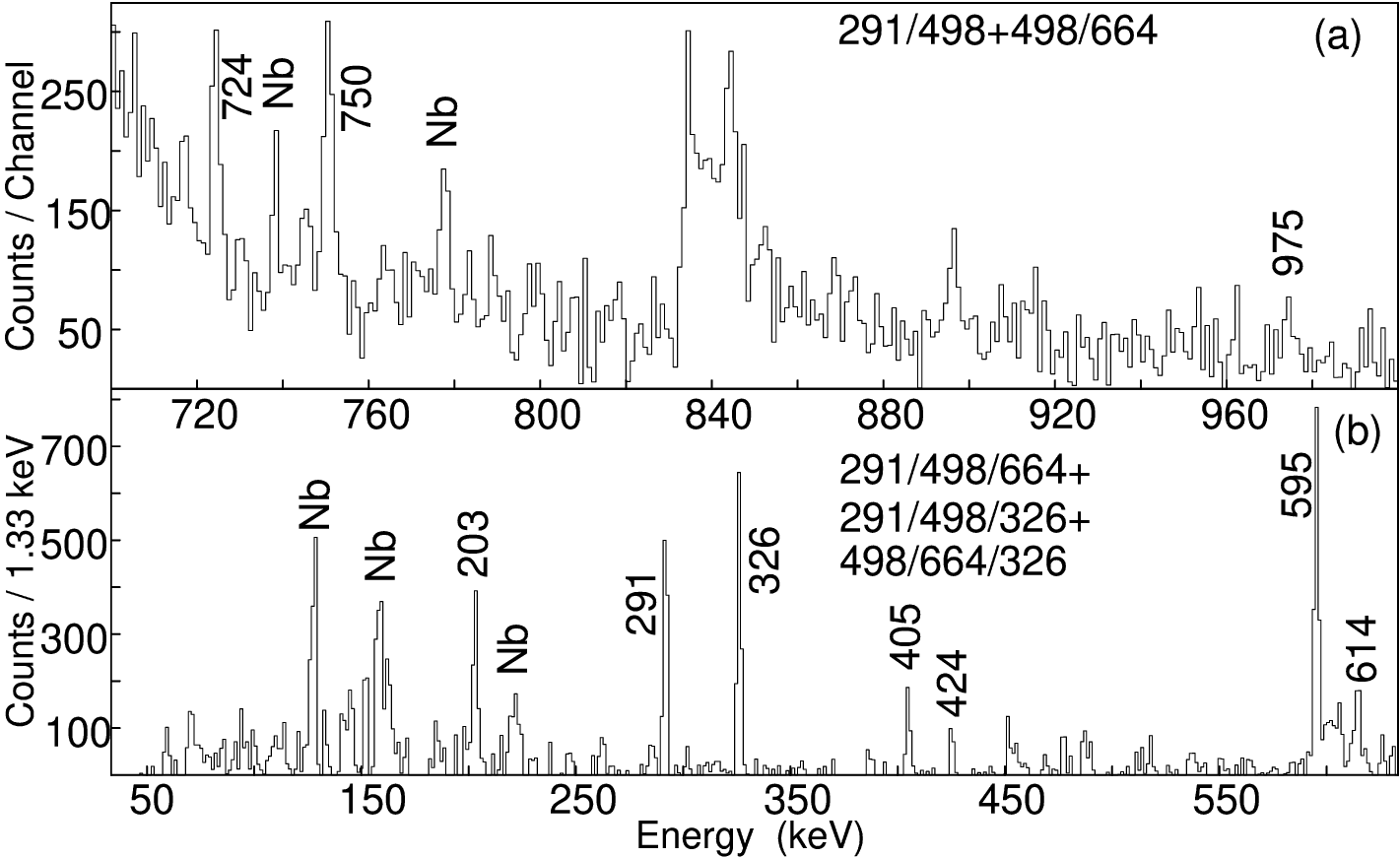}
 \caption{Coincidence spectra corresponding to the (a) added double gates of 291-498 and
   498-664, (b) added triple gates of 291-498-664, 291-498-326 and 498-664-326
   $\gamma$ rays in $^{143}$La, as obtained from the $^{252}$Cf fission data.}
 \label{143La-gate2-Cf}
\end{figure}

Using a sum gate of $291$ and $498$~keV  $\gamma$ - ray transitions
(\figurename~\ref{143La-gate-agata}), the previously reported transitions of $291$, $326$, $454$, $498$,
$518$, $536$, $570$, $595$, $615$, $641$ and $664$~keV can be clearly identified.
The transitions and levels in Bands (2), (4) and (5) in \figurename~\ref{LS_143La}
were assigned to  $^{142}$La in Ref.~\cite{Luo2021}.
 These Bands in \figurename~\ref{LS_143La} are now assigned to $^{143}$La
 in the present work, using the A and Z gated isotopically identified $\gamma$ - rays.
 It is clear from \figurename~\ref{Singles}, that, the $425$, $549$, $569$ and $705$~keV
 transitions are seen in the A and Z gated spectrum corresponding to $^{143}$La,
 but not in the corresponding spectrum for $^{142}$La.
 Furthermore, in \figurename~\ref{143La-gate1-Cf}(a) the triple gate of 
 $291$/$498$/$705$~keV transitions, shows the $205$~keV linking transition
 from Band (2) to Band (1). In \figurename~\ref{143La-gate1-Cf}(b),
 using triple gates of the $425$/$569$/$326$~keV transitions,
 the $459$~keV linking transition from Band (1) to Band (2) can be seen.
 The strong $595$~keV transition in Band (6) is also seen in \figurename~\ref {143La-gate2-Cf}.
 These coincidence spectra  from $^{252}$Cf spontaneous fission data assign
 these $\gamma$-rays to be the linking transitions from
 Bands (1) and (2). This also confirms that these transitions are indeed in the same nucleus
 and not two different nuclei, as assigned in Ref.~\cite{Wang2007,Luo2009,Luo2021}.

 Spectra in \figurename~\ref{143La-gate2-Cf} show the new transitions in $^{143}$La,
 based on  coincidences from $^{252}$Cf spontaneous fission data. In part (a), a
 summation of two double gates on $291/498$ and $498/664$~keV transitions,
 shows new higher energy $724$, $750$, and $975$~keV transitions.
 In part (b), by adding $291/498/664$, $291/498/326$ and $498/664/326$~keV
 triple gates, the new $424$~keV transition, as well as
 the previously identified $203$, $291$, $326$, $405$ and $595$~keV transitions, can be seen.
  Thus, the unique identification of $\gamma$ rays in the case of a nucleus  with low yields and
  with no known $\gamma$ transitions, can be challenging using only the $\gamma$-ray
  coincidences and the cross-coincidence
 relationships with the complementary fission fragments. Hence, the combination of two
 very powerful techniques used in this work is the key for obtaining detailed spectroscopy of
 such hard-to-populate nuclei.

\subsubsection{Angular Correlation Measurements}

\begin{figure}[ht]
 \includegraphics[width=\columnwidth]{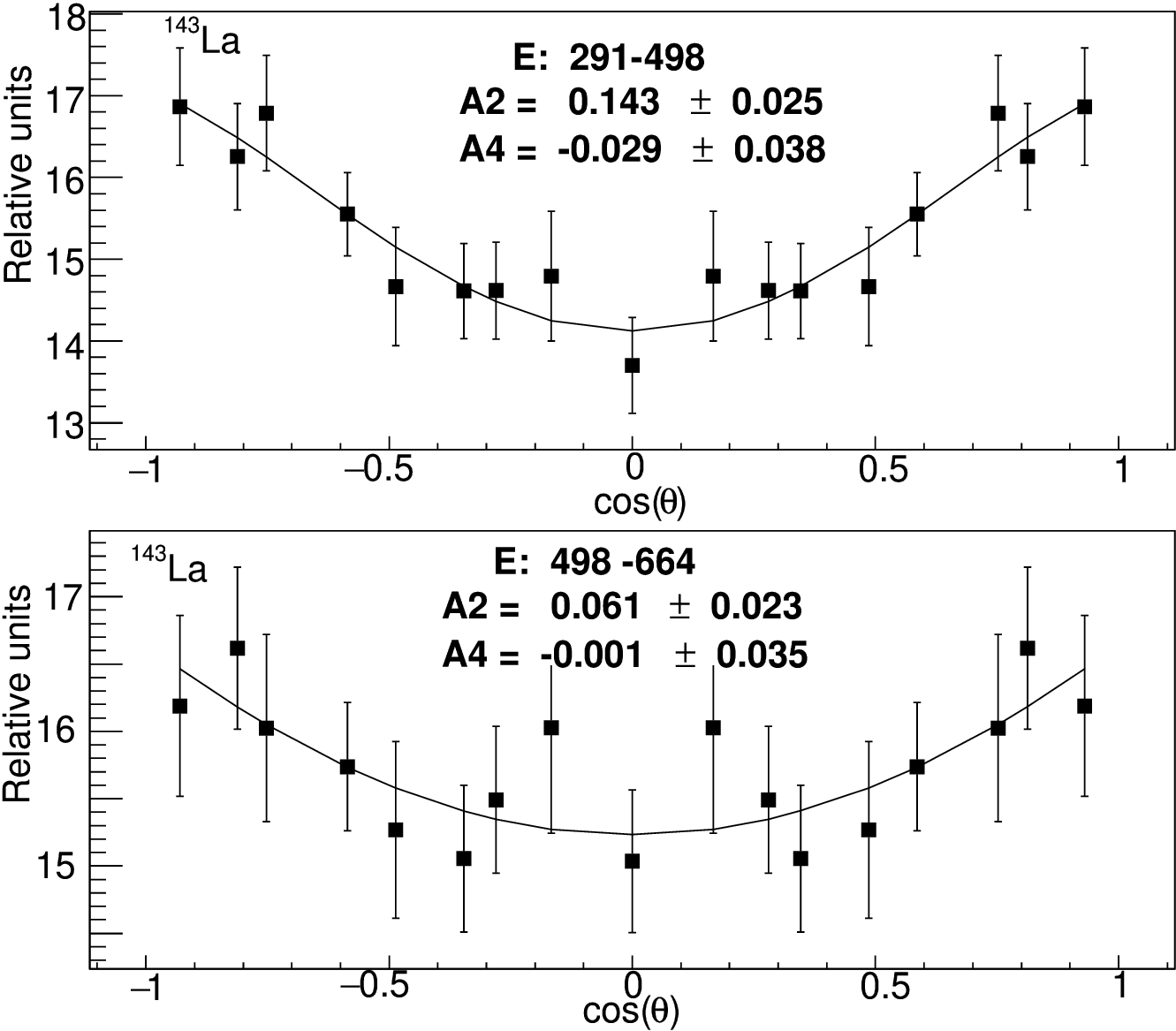}
  \caption{Angular correlation of the 291-498~keV cascade (upper panel)
    and the $498-664$~keV cascade (lower panel)
 in $^{143}$La, as obtained from the $^{252}$Cf spontaneous fission data.}
 \label{143La291-498}
\end{figure}

In the present work, the relatively large yields of $^{143}$La in $^{252}$Cf spontaneous
fission permitted the measurement of $\gamma$- ray angular correlations work, and the
results are shown in \figurename~\ref{143La291-498} and Table~\ref{tabangle_m}.
These are used to determine the J$^\pi$ of the states in $^{143}$La.
These newly measured angular correlations are combined with previous measurements for
the $\beta$-decay of $^{143}$Ba~\cite{Faller1988} to assign the J$^\pi$ of the
excited states of $^{143}$La.

\begin{table*}[ht]
  \centering
 \setlength{\extrarowheight}{5pt}
\caption{Angular correlations of cascades in $^{143}$La. Values from $^{143}$Ba $\beta$ decay in
  Ref.~\cite{Faller1988} are also included and labeled with an `$\ast$'. The spin assignments from the
  present work are marked with `$\ast\ast$'. All the cascades for the present angular
  correlations are chosen in such a way that one of the transitions is either E2 or E1 so
  that it has minimum (zero) mixing. The possible mixing ratio ($\delta$) for the other
  transition is quoted comparing the calculated (theoretical) angular correlation
  coefficients (A$_2$ and A$_4$) for different $\delta$ with the experimental values.}
	\begin{tabular*}{\linewidth}{@{\extracolsep{\fill}}ccccc}
	\hline\hline
	cascade (keV) & $A_2/A_4$ exp. & $A_2/A_4$ theo. & $\delta$(E2/M1) & possible decay pattern \\
        \hline
	719-291 & -0.23(4)/none$*$ & -0.14/0 & 0 & 5/2$^-$(E1)3/2$^+$(E2)7/2$^+$\\
        \hline
	& & &6.2(29) or 0.31(8) & 5/2$^-$(E1)5/2$^+$(E2/M1)7/2$^+$\\
        \hline
	&&& 0.44(30) & $**$5/2$^-$(E1)7/2$^+$(E2/M1)7/2$^+$\\
        \hline
	498-291 & 0.143(25)/-0.029(38) & (within 2$\sigma$) & -0.87(56)  & 9/2$^+$(E2)5/2$^+$(E2/M1)7/2$^+$\\
	\hline
        &&& 0.42(33) & $**$11/2$^+$(E2)7/2$^+$(E2/M1)7/2$^+$\\
	\hline
        719-261 & -0.14(3)/none$*$ & (within 2$\sigma$) & 0.78(54) & 5/2$^-$(E1)3/2$^+$(E2/M1)3/2$^+$\\
	\hline
        &&& -3.7(5) or 0.024(37) & 5/2$^-$(E1)5/2$^+$(E2/M1)3/2$^+$\\
	\hline
        && -0.15/0 & 0 &$**$5/2$^-$(E1)7/2$^+$(E2)3/2$^+$\\
        \hline
	498-664 & 0.061(23)/-0.001(35) & 0.10/0.01 (within 2$\sigma$) & & $**$15/2$^+$(E2)11/2$^+$(E2)7/2$^+$\\
	\hline
        326-664 & 0.025(43)/-0.013(65) & & 5.2(20) or 0.16(7) & $**$17/2$^+$(E2/M1)15/2$^+$(E2)11/2$^+$\\
	\hline
	\end{tabular*}
\label{tabangle_m}
\end{table*}

Results from the $\beta$-decay relevant to the level scheme from the present
work are briefly summarized below.  The ground state of $^{143}$La was assigned a
J$^\pi$=7/2$^+$ based on the logft value and systematics. The state at 30~keV was
assigned a J$^\pi$ 3/2$^+$ based on the E2 character of the transition to the ground state,
obtained from the measurement of internal conversion coefficients.
The state at $1010$~keV level was assigned J$^\pi$=5/2$^-$ based on the logft value
and the decay pattern. These assignments are adopted in the present work.
Foller {\it et al.}~\cite{Faller1988} further reported other angular correlation
measurements of transitions from low-spin non-yrast states (generally relatively poorly populated
in a fission reaction) that can be used to obtain the J$^\pi$ of states populated
in the present work. The angular correlation for the $719-291$~keV transitions
(719 keV transition from the 5/2$^-$ state at $1010$~keV populates the state at 291 keV,
which in turn feeds the 7/2$^+$ ground state) was used to determine the J$^\pi$ of the
state at $291$~keV. The $261$ and $291$~keV transitions (see \figurename~\ref{LS_143La})
were assigned as mixed (E2/M1) transition from the measured internal conversion coefficients.
Thus, the possible values of the J$^\pi$ of the state at $291$~keV are 3/2$^+$, 5/2$^+$ or
7/2$^+$. It should be noted that for the $261$~keV transition from the state at $291$~keV
to the 3/2$^+$ state at $30$~keV level, the measured $\alpha_k$ value is in better
agreement with the calculated value for an E2 transition. For an assignment of 3/2$^+$ spin
for the state at $291$~keV in the cascade 5/2$^-$ ($719, E1)$ 3/2$^+$ ($291, E2)$ 7/2$^+$,
the mixing ratio ($\delta$) is very small. The calculated angular correlation
coefficients for $\delta$=0, do not match the experimental values given in
Table~\ref{tabangle_m}, and thus we can discard the assignment of 3/2$^+$ spin for the
state at $291$~keV. Among the other two possibilities (5/2$^+$ or 7/2$^+$) the calculations
match the experimental angular correlation coefficients for different mixing ratios of
$291$~keV transition. The values of the mixing ratio ($\delta$) of the $291$~keV transition
are found to be $6.2$($29$) or $0.31$($8$) in the case of J$^\pi$= 5/2$^+$ and $0.44$
if J$^\pi$=7/2$^+$.

In Table~\ref{tabangle_m} the measured angular correlation of the $498-291$~keV cascade (Band 1)
is compared with the theoretical values assuming a pure E2 character (zero mixing) for the
$498$~keV transition. For an assignment of J$^\pi$ of 5/2$^+$ for the level at $291$~keV,
the experimental values of angular correlation coefficients are within 2$\sigma$ of the
calculated values for a $\delta$ = $-0.87$($56$). On the other hand, for an assignment of
J$^\pi$ of 7/2$^+$, the experimental values are well within the limit of theoretical values
for $\delta$= $0.42$($33$). The angular correlation coefficients for $719-291$~keV cascade,
reported in~\cite{Faller1988} and that for $498-291$~keV cascade from the present work,
show a consistent value of $\delta$ of $0.44$($30$) and $0.42$($33$) respectively
(Table~\ref{tabangle_m}), for the $291$~keV transition only if the $291$~keV level has a
spin-parity of 7/2$^+$. The assignment of 7/2$^+$ to this level is further validated by the
angular correlation of the $719-261$~keV cascade as shown in Table~\ref{tabangle_m}.
Among the three possible spin sequences, the theoretical angular correlation coefficients
($-0.15/0$) for the E1-E2 (both $\delta$ =0) 5/2$^-$ $719$ (E1) 7/2$^+$ $261$ (E2) 3/2$^+$ cascade.
The 7/2$^+$ assignment also corroborates with the E2 nature of the $261$~keV transition,
decaying from the same state. Based on the above angular correlation analysis,
the spin-parity of the $291$~keV level is unambiguously shown to be 7/2$^+$. 

The angular correlation, shown in the lower panel of \figurename~\ref{143La291-498}, for the
$498-664$ cascade (Band 1), is consistent with a pure E2-E2 cascade. As the state at
$789$~keV was not observed in $\beta$-decay~\cite{Faller1988}, it indicates its high spin character.
The E2 character of the $498$ and $664$~keV transitions are also in line with those in neighbouring
nuclei (discussed in the next section).  The angular correlation of the $326-664$~cascade
(Band 6 and Band 1) was also measured in the present work and was found to have a positive
A2 value. Assuming the $664$~keV transition to be of an E2 character, the positive A2 value
yields a mixed E2/M1 character for the $326$~keV transition. This indicates that Band (3)
has a positive parity as assigned in Ref.~\cite{Luo2009}.

\section{Discussion}

\begin{figure}[h!]
\includegraphics[width=\columnwidth,clip=true]{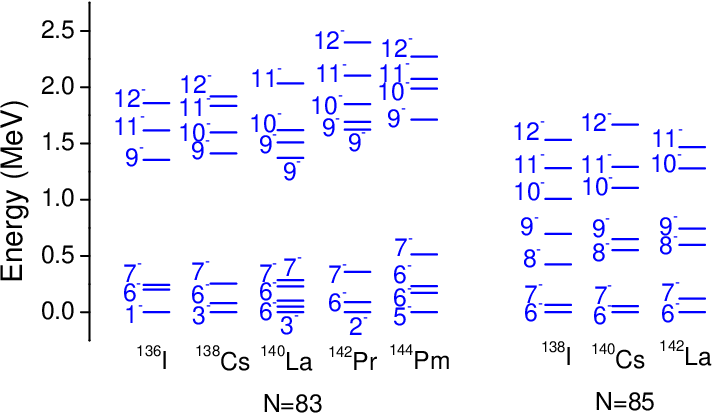}
\caption{Systematics of the excited states for even-A $N=83$ and $N=85$ isotones.
  For the $N=83$ isotones, the energies correspond to the  excited states.
  For $N=85$ isotones, the energies are with respect to the lowest 6$^-$ level.}
 \label{EvenLa-com}
\end{figure}

\begin{figure}[ht]
 \includegraphics[width=0.8\columnwidth,clip=true]{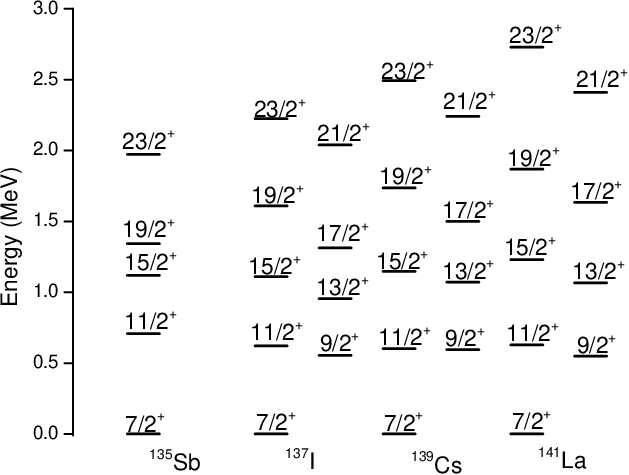}
 \caption{Systematics of positive parity excited states (with respect to
   7/2$^+$ state) for odd-A $N=84$ isotones.}
 \label{141La-com}
\end{figure}

\begin{figure}[ht]
 \includegraphics[width=0.8\columnwidth,clip=true]{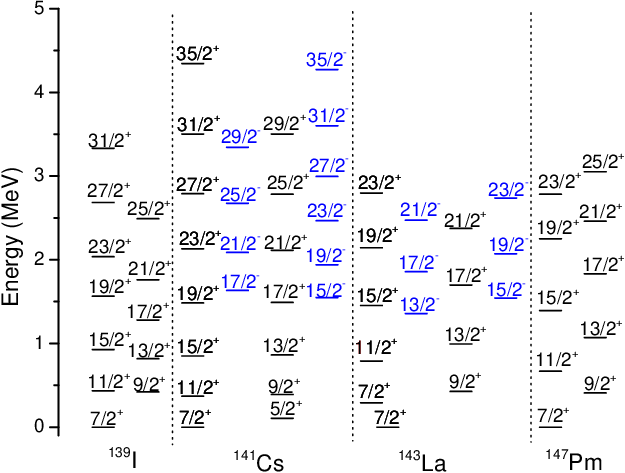}
 \caption{Systematics of excited states for odd-A $N=86$ isotones.
 For $^{143}$La, only the states corresponding to the parity doublet structures are shown. 
 The positive and negative parity states are shown in black and blue colour respectively. }
 \label{143La-com}
\end{figure}

The systematics for the excited state for the $N=83$ to $86$ of
La and neighboring isotones are shown in Figs.~\ref{EvenLa-com},~\ref{141La-com} and~\ref{143La-com}.
Although the present induced fission data can only determine the excited energy levels from the
transition coincidence relationships and their intensities, the energy systematics
of isotones and the following discussed shell model calculations support the tentative spin/parity
assignments.


From the excitation energy pattern of the $N=83$ isotones (\figurename~\ref{EvenLa-com}),
it can be seen that the negative parity excited states identified in the present
work for $^{140}$La follow well the systematics of the neighbouring isotones
for $Z=53$ to $Z=61$ nuclei.
The level sequence and spacing for these isotopes exhibit a strong similarity.
The excitations up to spin $10^-$ states are almost constant from $^{136}$I ($Z=53$)
to $^{140}$La ($Z=57$).
For $^{142}$Pr ($Z=59$) and $^{144}$Pm ($Z=61$), the higher
spin states shift to relatively higher excitations. This could  be due to the increasing
collectivity related to the larger number of valence protons outside the $Z=50$ major shell closure. 

The systematics of $N=85$ isotones for I ($Z=53$), Cs ($Z=55$) and La ($Z=57$)
are also plotted in \figurename~\ref{EvenLa-com}. Very limited data
for $^{144}$Pr~\cite{Wang2015} is available and hence is not shown in the figure.
A remarkable similarity is also observed for
the exited states in $N=85$ $^{138}$I,$^{140}$Cs, $^{142}$La. 
These systematics were used  in the tentative assignment of the spin-parity of the
new high spin states observed in $^{142}$La in the present work. 

In the case of $N=84$ isotones (\figurename~\ref{141La-com}), the yrast cascade patterns
of all these isotopes from Sb ($Z=51$) to La ($Z=57$) are also very similar.
The first excited state reported in these isotones is $5/2^+$, the corresponding
state is not observed in $^{141}$La from the present work
and therefore not shown in \figurename~\ref{141La-com}.
The other Band built on the $1218$~keV level,
observed in $^{141}$La, decays to the yrast Band at $628$~keV (11/2$^+$).
This Band is not yet observed in the $N=84$ isotones of the
other lower $Z$ nuclei and therefore not shown in \figurename~\ref{141La-com}.

The systematics of the excited states of N=86 isotones are shown in \figurename~\ref{143La-com}.
The parity doublet structure observed in $^{143}$La in the present work is very
similar to the parity doublet structure reported in neighbouring isotope $^{141}$Cs.
In the case of I and Pm isotopes having the same neutron number, only positive parity bands
are reported. 


\subsection{Shell Model Calculations}

The shell model  calculations presented here have been performed in the
model space of $50 \leq Z \leq 82 $ and $82 \leq N \leq 126 $.
The  effective Hamiltonian is based on the complete effective nuclear force
V$_{\text{MU}}$+LS~\cite{otsuka2010}, which consists of the Gaussian central force,
the $\pi + \rho$ meson-exchange tensor force, and M3Y spin-orbit force~\cite{bertsch1977}.
V$_{\text{MU}}$+LS has been successfully employed in the $psd$~\cite{yuan2012},
$sdpf$~\cite{utsuno2012}, $pfsdg$ regions~\cite{togashi2015}, and the nearby
regions of $^{132}$Sn~\cite{yuan2016PLB} and $^{208}$Pb~\cite{yuan_pb208,zhangzy2021,yanghb2022}.
Recently, V$_{\text{MU}}$+LS was employed to study the medium-heavy nuclei around $^{132}$Sn
and  $^{208}$Pb~\cite{yuan2020,review} in a unified way. Specifically, it has well
reproduced the separation and excitation energies, as well as the nuclear level densities
of nuclei with $50 \leq Z \leq 56 $ and $80 \leq N \leq 84 $~\cite{Chen2023}.
All the shell model calculations discussed here are performed with the
code KSHELL~\cite{shimizu2013}.

\begin{figure}[ht]
\includegraphics[width=0.9\columnwidth,clip=true]{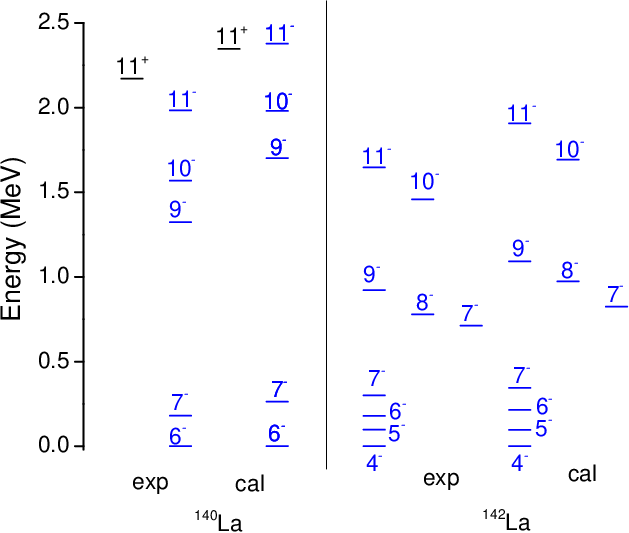}
\caption{Comparison of experimental energies  (exp) of the excited states of even-$A$,
  $^{140, 142}$La isotopes with the Shell Model calculations (cal).
  The positive and negative parity states are shown in black and blue colour respectively. 
}
\label{EvenLa-cal}
\end{figure}

\begin{figure}[t]
 \includegraphics[width=0.8\columnwidth,clip=true]{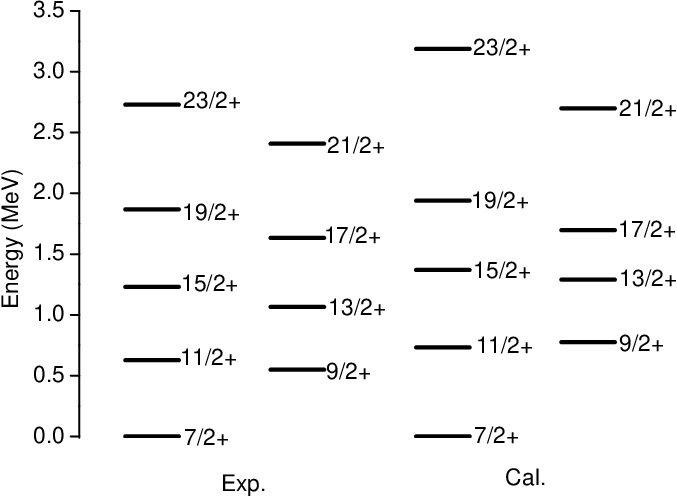}
 \caption{Comparison of experimental energies  (Exp.) of the excited states of $^{141}$La
   with Shell Model calculations (Cal.). }
 \label{141La-cal}
\end{figure}
 
 In the case of $^{142}$La, the calculations are restricted to the condition that a maximum
 of two protons are allowed to occupy the $\pi1d_{3/2}$, $\pi0h_{11/2}$, and $\pi2s_{1/2}$ orbital.
 In the calculations for  $^{143}$La, the restrictions for the  proton excitation are the
 same as in  $^{142}$La. Additionally, the number of neutrons particles in
the  $\nu$f$_{7/2}$ orbital can range between two and four.
These constraints for $^{142,143}$La arise from computational limitations due to the
very large model space as the number of valence neutrons increases above $N=82$ closed shell.
The calculated results for the excited states of even-$A$ La
isotopes are compared with the corresponding experimental levels and are shown in Figs.~\ref{EvenLa-cal}.
In this figure, only states lying above  $6^{-}$ and $4^{-}$, respectively, are shown, as the
low-lying levels below $6^{-}$ in $^{140}$La and below $4^{-}$ in $^{142}$La, are not observed
in the present work.


\begin{table}[!htbp]
	\centering
	\caption{\label{configuration_with_ex_o-o} The two leading components of the Shell model configurations for $^{140,142}$La.The calculated excitation energies are shown in column Ex
		and are in MeV. The columns $g7$, $d5$, $h11$, $h9$, $f7$, and $p3$ represent the occupancy in the $\pi0g_{7/2}$, $\pi 1d_{5/2}$, $\pi 0h_{11/2}$ , $\pi 0h_{9/2}$, $\nu 1f_{7/2}$, and $\nu 2p_{3/2}$ orbitals, respectively. The last column shows the percentage of each configuration.}
	\setlength{\tabcolsep}{3pt}
	\begin{tabular}{ccccccccccc}
		\hline\hline
		Nuclide & $J^{\pi} $  &  Ex  & $g7$ & $d5$ & $h11$  & $h9$  & $f7$  & $p3$  & \% \\\hline
		
		$^{140}$La &   $2^{-}_{1}$  &  0      &  4  &  3  &  0  &  0  &  1  &  0  &  34   \\
		&                &         &  6  &  1  &  0  &  0  &  1  &  0  &  17   \\
		&   $2^{-}_{2}$  &  0.077  &  5  &  2  &  0  &  0  &  1  &  0  &  31   \\
		&                &         &  3  &  4  &  0  &  0  &  1  &  0  &  13   \\
		&   $6^{-}_{1}$  &  0.129  &  5  &  2  &  0  &  0  &  1  &  0  &  37   \\ 
		&                &         &  3  &  4  &  0  &  0  &  1  &  0  &  11   \\
		&   $4^{-}_{1}$  &  0.137  &  5  &  2  &  0  &  0  &  1  &  0  &  36   \\ 
		&                &         &  3  &  4  &  0  &  0  &  1  &  0  &  12   \\
		&   $3^{-}_{1}$  &  0.166  &  5  &  2  &  0  &  0  &  1  &  0  &  35   \\ 
		&                &         &  3  &  4  &  0  &  0  &  1  &  0  &  11   \\
		&   $1^{-}_{1}$  &  0.221  &  5  &  2  &  0  &  0  &  1  &  0  &  17   \\ 
		&                &         &  6  &  1  &  0  &  0  &  1  &  0  &  15   \\
		&   $5^{-}_{1}$  &  0.221  &  5  &  2  &  0  &  0  &  1  &  0  &  36   \\ 
		&                &         &  3  &  4  &  0  &  0  &  1  &  0  &  10   \\
		&   $0^{-}_{1}$  &  0.280  &  5  &  2  &  0  &  0  &  1  &  0  &  28   \\ 
		&                &         &  3  &  4  &  0  &  0  &  1  &  0  &  19   \\
		&   $7^{-}_{1}$  &  0.392  &  5  &  2  &  0  &  0  &  1  &  0  &  33   \\
		&                &         &  3  &  4  &  0  &  0  &  1  &  0  &  14   \\
		&   $8^{-}_{1}$  &  1.578  &  5  &  2  &  0  &  0  &  1  &  0  &  54   \\
		&                &         &  3  &  4  &  0  &  0  &  1  &  0  &   7    \\
		&   $9^{-}_{1}$  &  1.830  &  5  &  2  &  0  &  0  &  1  &  0  &  45   \\
		&                &         &  3  &  4  &  0  &  0  &  1  &  0  &  12   \\
		&   $8^{+}_{1}$  &  1.839  &  4  &  2  &  1  &  0  &  1  &  0  &  35   \\
		&                &         &  6  &  0  &  1  &  0  &  1  &  0  &  22   \\
		&   $10^{-}_{1}$ &  2.110  &  6  &  1  &  0  &  0  &  1  &  0  &  42   \\
		&                &         &  4  &  3  &  0  &  0  &  1  &  0  &  29   \\
		&   $11^{+}_{1}$ &  2.476  &  4  &  2  &  1  &  0  &  1  &  0  &  45   \\
		&                &         &  6  &  0  &  1  &  0  &  1  &  0  &  15   \\
		&   $11^{-}_{1}$ &  2.506  &  4  &  3  &  0  &  0  &  1  &  0  &  35   \\
		&                &         &  6  &  1  &  0  &  0  &  1  &  0  &  33   \\
		\hline
		
		$^{142}$La &   $0^{-}_{1}$  &  0      &  4  &  3  &  0  &  0  &  3  &  0  &  13   \\
		&                &         &  4  &  3  &  0  &  0  &  2  &  1  &  11    \\
		&   $2^{-}_{1}$  &  0.180  &  4  &  3  &  0  &  0  &  3  &  0  &  17   \\
		&                &         &  4  &  3  &  0  &  0  &  2  &  1  &  12    \\
		&   $1^{-}_{1}$  &  0.254  &  4  &  3  &  0  &  0  &  3  &  0  &  19   \\
		&                &         &  4  &  3  &  0  &  0  &  2  &  1  &  10    \\
		&   $4^{-}_{1}$  &  0.387  &  4  &  3  &  0  &  0  &  3  &  0  &  14   \\
		&                &         &  4  &  3  &  0  &  0  &  2  &  1  &   8    \\
		&   $4^{-}_{1}$  &  0.417  &  4  &  3  &  0  &  0  &  3  &  0  &  10   \\
		&                &         &  5  &  2  &  0  &  0  &  3  &  0  &   8    \\
		&   $5^{-}_{1}$  &  0.484  &  3  &  4  &  0  &  0  &  3  &  0  &  10    \\
		&                &         &  5  &  2  &  0  &  0  &  3  &  0  &   9    \\
		&   $6^{-}_{1}$  &  0.601  &  5  &  2  &  0  &  0  &  3  &  0  &  15   \\
		&                &         &  3  &  4  &  0  &  0  &  3  &  0  &   9    \\
		&   $7^{-}_{1}$  &  0.731  &  3  &  4  &  0  &  0  &  3  &  0  &  16   \\
		&                &         &  5  &  2  &  0  &  0  &  3  &  0  &  15   \\
		&   $7^{-}_{2}$  &  1.211  &  4  &  3  &  0  &  0  &  3  &  0  &  19  \\
		&                &         &  4  &  3  &  0  &  0  &  2  &  1  &  11 \\ 
		&   $8^{-}_{1}$  &  1.361  &  4  &  3  &  0  &  0  &  3  &  0  &  14   \\
		&                &         &  5  &  2  &  0  &  0  &  3  &  0  &   6    \\
		&   $9^{-}_{1}$  &  1.479  &  3  &  4  &  0  &  0  &  3  &  0  &  17   \\
		&                &         &  5  &  2  &  0  &  0  &  3  &  0  &  12   \\
		&   $10^{-}_{1}$ &  2.080  &  4  &  3  &  0  &  0  &  3  &  0  &  20   \\
		&                &         &  4  &  3  &  0  &  0  &  2  &  1  &   7    \\
		&   $11^{-}_{1}$ &  2.294  &  3  &  4  &  0  &  0  &  3  &  0  &  20   \\
		&                &         &  5  &  2  &  0  &  0  &  3  &  0  &  11   \\
		\hline
	\end{tabular}
\end{table}

\begin{table}[!htbp]
\centering
\caption{\label{configuration_with_ex_o-e}
  Same as in Table II, for $^{141,143}$La.
}
\setlength{\tabcolsep}{3pt}
\begin{tabular}{ccccccccccc}
\hline\hline
Nuclide & $J^{\pi} $  &  Ex  & $g7$ & $d5$ & $h11$  & $h9$  & $f7$  & $p3$  & \% \\\hline

$^{141}$La &   $7/2^{+}_{1}$  & 0     &  5 &  2  &  0  &  0  &  2  &  0  &  20   \\
           &                  &       &  3  &  4  &  0  &  0  &  2  &  0  &  11   \\
           &   $11/2^{+}_{1}$ & 0.732 &  5  &  2  &  0  &  0  &  2  &  0  &  19   \\
           &                  &       &  3  &  4  &  0  &  0  &  2  &  0  &  11   \\
           &   $9/2^{+}_{1}$  & 0.775 &  5  &  2  &  0  &  0  &  2  &  0  &  16   \\
           &                  &       &  5  &  2  &  0  &  0  &  1  &  1  &   7    \\
           &   $13/2^{+}_{1}$ & 1.289 &  5  &  2  &  0  &  0  &  2  &  0  &  16   \\
           &                  &       &  3  &  4  &  0  &  0  &  2  &  0  &  7    \\
           &   $15/2^{+}_{1}$ & 1.370 &  5  &  2  &  0  &  0  &  2  &  0  &  20   \\
           &                  &       &  3  &  4  &  0  &  0  &  2  &  0  &  11   \\
           &   $17/2^{+}_{1}$ & 1.697 &  5  &  2  &  0  &  0  &  2  &  0  &  23   \\
           &                  &       &  3  &  4  &  0  &  0  &  2  &  0  &  12   \\
           &   $19/2^{+}_{1}$ & 1.939 &  5  &  2  &  0  &  0  &  2  &  0  &  22   \\
           &                  &       &  3  &  4  &  0  &  0  &  2  &  0  &  16   \\
           &   $21/2^{+}_{1}$ & 2.698 &  5  &  2  &  0  &  1  &  1  &  0  &  23   \\
           &                  &       &  3  &  2  &  2  &  1  &  1  &  0  &  10    \\
           &   $23/2^{+}_{1}$ & 3.188 &  5  &  2  &  0  &  1  &  1  &  0  &  15   \\
           &                  &       &  3  &  4  &  0  &  1  &  1  &  0  &   8    \\
\hline

$^{143}$La &   $7/2^{+}_{1}$ & 0.     &  3  &  4  &  0  &  0  &  4  &  0  &  14  \\
           &                 &        &  5  &  2  &  0  &  0  &  4  &  0  &   7   \\
           &   $5/2^{+}_{1}$ & 0.021 &  4  &  3  &  0  &  0  &   4  &  0  &  10   \\
           &                  &       & 3  &  4  &  0  &  0  &   4  &  0  &   8   \\
           &   $3/2^{+}_{1}$  & 0.050 & 4  &  3  &  0  &  0  &   4  &  0  &  16   \\
           &                  &       & 4  &  3  &  0  &  0  &   3  &  1  &   7    \\
           &   $5/2^{+}_{2}$ & 0.140 &  4  &  3  &  0  &  0  &   4  &  0  &  13   \\
           &                  &       &  3  &  4  &  0  &  0  &  4  &  0  &   5    \\
           &   $7/2^{+}_{1}$ & 0.406  &  4  &  3  &  0  &  0  &  4  &  0  &  19   \\
           &                  &       &  4  &  3  &  0  &  0  &  3  &  1  &  11   \\
           &   $11/2^{+}_{1}$ & 0.466 &  3  &  4  &  0  &  0  &  4  &  0  &  13   \\
           &                  &       &  3  &  4  &  0  &  0  &  3  &  1  &   6   \\
           &   $9/2^{+}_{1}$ & 0.478 &  3  &  4  &  0  &  0  &  4  &  0  &   8   \\
           &                  &       &  4  &  3  &  0  &  0  &  4  &  0  &   7   \\ 
           &   $5/2^{+}_{3}$ & 0.542 &  5  &  2  &  0  &  0  &  4  &  0  &  16   \\
           &                  &       & 5  &  2  &  0  &  0  &  3  &  1  &  6   \\
           &   $7/2^{+}_{3}$ & 0.632 &  5  &  2  &  0  &  0  &  4  &  0  &  16   \\
           &                  &       & 3  &  4  &  0  &  0  &  4  &  0  &   7   \\
           &   $9/2^{+}_{2}$ & 0.634 &  4  &  3  &  0  &  0  &  2  &  0  &  14   \\
           &                  &       &  3  &  4  &  0  &  0  & 3  &  1  &   7   \\ 
           &   $9/2^{+}_{3}$ & 0.855 &  3  &  4  &  0  &  0  &  4  &  0  &  7   \\
           &                  &       &  5  &  2  &  0  &  0  &  4  &  0  &  7   \\
           &   $11/2^{+}_{2}$ & 0.886 &  4  &  3  &  0  &  0  &  2  &  0  &  18   \\
           &                  &       &  4  &  3  &  0  &  0  &  3  &  1  &  11   \\
           &   $13/2^{+}_{1}$ & 0.974 &  3  &  4  &  0  &  0  &  4  &  0  &   8   \\
           &                  &       &  3  &  4  &  0  &  0  &  3  &  1  &   8   \\ 
           &   $15/2^{+}_{1}$ & 1.022 &  3  &  4  &  0  &  0  &  4  &  0  &  13   \\
           &                  &       &  3  &  4  &  0  &  0  &  3  &  1  &   8   \\
           &   $11/2^{+}_{3}$ & 1.131 &  5  &  2  &  0  &  0  &  4  &  0  &  18   \\
           &                  &       &  3  &  4  &  0  &  0  &  3  &  1  &   7   \\
           &   $13/2^{-}_{1}$ & 2.62 &   3 &  3  &  0  &  1  &  4 &  0  &  18  \\
           &                  &       &  3  &  2  &  0  &  1  &  2  &  0  &  16   \\ 
\hline
\end{tabular}
\end{table}

\begin{figure}[t]
	\includegraphics[width=\columnwidth]{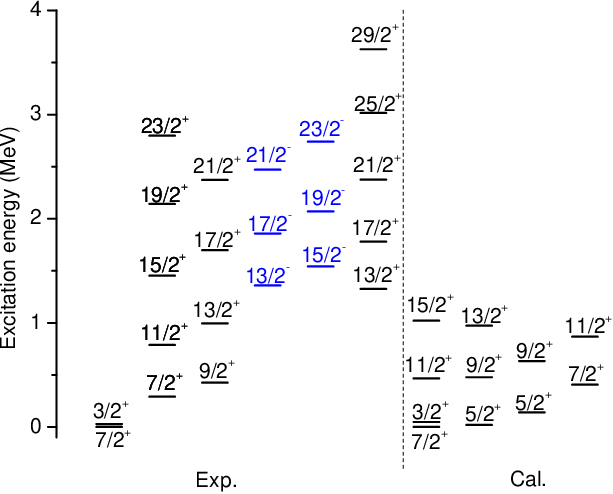}
	\caption{Comparison of experimental energies  (Exp.) of the excited states of $^{143}$La with Shell Model calculations (Cal.). }
	\label{143La-cal}
\end{figure}



As can be seen from \figurename~\ref{EvenLa-cal}, the calculated excitation energies in
$^{140-142}$La are consistent with the tentative assignment of spin and parity of the
newly observed states. For the odd-odd nuclei $^{140,142}$La, the negative parity states can be divided into two groups based on their dominant shell model configurations. These configurations for the various states are given in Table~\ref{configuration_with_ex_o-o}. For the lower
group of states, including the $6^{-}$, $7^{-}$, and $9^{-}$ states in $^{140}$La and the $5^{-}$, $6^{-}$, $7^{-}_{1}$, and $9^{-}$ states in $^{142}$La, are dominated by the $\pi (0\emph{g}_{7/2})^{5}(1\emph{d}_{5/2})^{2}\nu (1\emph{f}_{7/2})^{n}$ and $\pi(0\emph{g}_{7/2})^{3}(1\emph{d}_{5/2})^{4}\nu(1\emph{f}_{7/2})^{n}$ configuration, with $n$ representing
the valence neutron number.
For the others, the $4^{-}$, $10^{-}$, and $11^{-}$ states in $^{140}$La are dominated by the $\pi(0\emph{g}_{7/2})^{4}(1\emph{d}_{5/2})^{3}\nu(1\emph{f}_{7/2})$ and the $\pi(0\emph{g}_{7/2})^{6}(1\emph{d}_{5/2})\nu(1\emph{f}_{7/2})$ configurations, while 
the $\pi(0\emph{g}_{7/2})^{4}(1\emph{d}_{5/2})^{3}\nu(1\emph{f}_{7/2})^{3}$ is the leading component in the $7^{-}_{2}$, $8^{-}$,and $10^{-}$ states in $^{142}$La. Overall, the dominant configurations represent a larger fraction in $^{140}$La as compared to $^{142}$La, reflecting an increase in collectivity with increasing number of valence neutrons. Regarding the positive parity states in $^{140}$La, the $8^{+}$ and $11^{+}$ states are dominated ($\approx$40\%) by the $\pi(0\emph{g}_{7/2})^{4}(1\emph{d}_{5/2})^{2}(0\emph{h}_{11/2})\nu(1\emph{f}_{7/2})$ configuration.
However, only the $11^{+}$ state is tentatively identified in the present experiment,
while the $8^{+}$ state could not be observed.

For $^{141}$La, the calculated energies align fairly well with the observed levels as shown in FIG.~\ref{141La-cal}. This agreement is expected, given that single-particle character tends to be more pronounced in odd-\emph{A} nuclei than in odd-odd nuclei.
Shell model configurations for various states in $^{141-143}$La are given in Table~\ref{configuration_with_ex_o-e}. The most probable configuration of the first $7/2^{+}$, $9/2^{+}$, $11/2^{+}$, $13/2^{+}$, $15/2^{+}$, $17/2^{+}$, and $19/2^{+}$ states is $\pi(0\emph{g}_{7/2})^{5}(1\emph{d}_{5/2})^{2}\nu(1\emph{f}_{7/2})^{2}$. On this basis, the $21/2^{+}$ and $23/2^{+}$ states result from a particle-hole excitation from $\nu1\emph{f}_{7/2}$ to $\nu1\emph{f}_{5/2}$. The intermediate values of the fraction of dominant contribution can be seen to lie between those for $^{140}$La and $^{142}$La. The new band built on the 1218 keV state in $^{141}$La has not been predicted in the present shell-model calculation.
As mentioned earlier in the case of $^{143}$La, the restricted large valence space and very large computing time do not allow for a comparison of all of the measured  states. The major contributions only for a few low-lying  states are shown in Table~\ref{configuration_with_ex_o-e}. A comparison with the shell model calculations for a restricted number of states for $^{143}$La is shown in Fig.~\ref{143La-cal}.



\subsection{Possible octupole deformation in $^{143}$La}

\begin{figure}[ht]
  	\includegraphics[width=0.75\columnwidth]{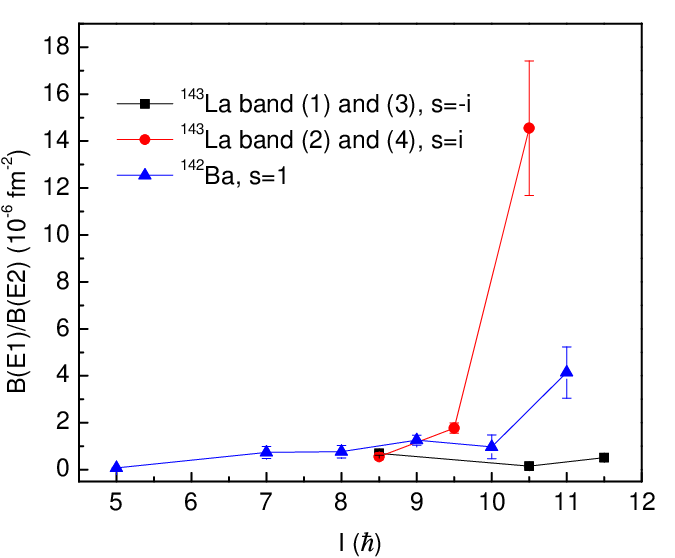}
	\caption{ B(E1)/B(E2) ratios of the $s=\pm i$ band structures (namely Bands (1), (3) and Bands (2), (4)) in $^{143}$La.}
	\label{143La-BE1E2}
\end{figure}

\begin{figure}[ht]
  \includegraphics[width=\columnwidth]{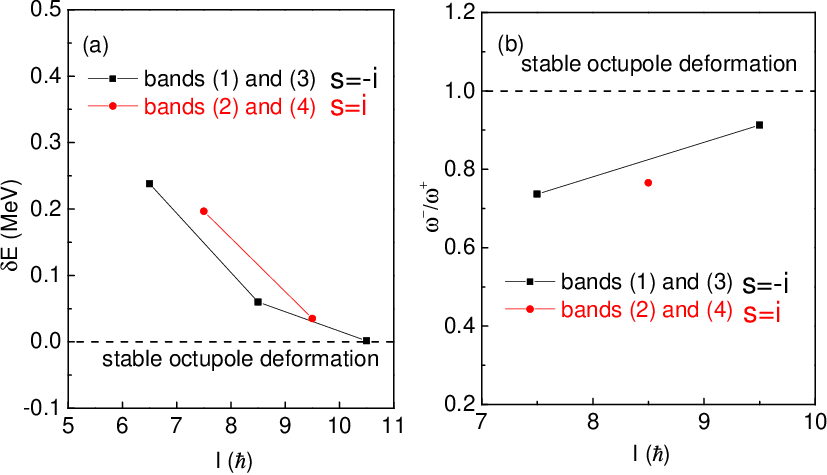}
	\caption{ The energy displacement $\delta E$ (a) and the rotational frequency ratio $\omega ^-/\omega^+$ (b) of the $s=\pm i$ band structures (namely, Bands (2), (4) and (1), (3)) in $^{143}$La.}
	\label{143La-DE-frequency-ratio}
\end{figure}

As the high spin level structure of $^{143}$La shows pairs of opposite parity bands,
with connecting $E1$ transitions between the two, the excited structures of $^{143}$La are further
discussed in the context of the possible presence of reflection asymmetric shape resulting in octupole deformation. 
 

Bands (1) and (3) in $^{143}$La were assigned as $s=i$ octupole doublet by
Ref.~\cite{Luo2009} with a 5/2$^+$ band-head. With the new assignment of the band-head
from the current work, these two bands should now form the $s=-i$ octupole doublet,
as the spins of the levels of the two bands increment by 1$\hbar$ in the new assignments.
The negative parity Band (3) (labeled as Band (1) in Ref.~\cite{Luo2009}) was interpreted as the single particle levels coupling to
the 1$^-$ and 3$^-$ levels in the $^{142}$Ba core \cite{Luo2009}.
Taking this interpretation to Bands (2) and (4), Band (4) can form the negative-parity
part of the $s=i$ doublet with Band (2).
It should be noted that multiple octupole doublets in a single nucleus have been seen only in limited cases.
Recently, three octupole doublets were reported in the neighboring $^{145}$Ba nucleus \cite{Bre2023}.

Energy displacement $\delta E$ between the opposite parity states of the same spin
and rotational frequency ratios $\omega ^\mp/\omega ^\pm$ are quantities to evaluate
the stability of octupole deformation~\cite{Naz85}. These quantities are calculated as follows:

\begin{equation}
\delta E(I)=E(I)\mp-\frac{E(I+1)^\pm-E(I-1)^\pm}{2}
\end{equation} 

\begin{equation}
\frac{\omega (I)\mp}{\omega (I)\pm}=2\times \frac{E(I+1)^\mp-E(I-1)^\mp}{E(I+2)^\pm-E(I-2)^\pm}
\end{equation}

B(E1)/B(E2) ratios of the octupole doublet are another such an observable.

We first look at the B(E1)/B(E2) ratios of the the octupole doublet.
The B(E1)/B(E2) ratios of the the octupole doublet with simplex quantum number $s=\pm i$
are extracted from the measured intensities of $\gamma$ rays and
are depicted in \figurename~\ref{143La-BE1E2}.
The ratios of the new $s=i$ band structures show an increasing trend with spin and have a much
larger value at higher spin.
The reason for this is unclear and could be related to additional uncertainties
in the extracted intensities beyond the fitting errors.
The sharply increasing B(E1)/B(E2) ratios for the $s=i$ structure were also reported in $^{143}$Ba,
with large staggering at higher spins~\cite{Bre2023}.
The current data on $^{143}$La do not extend to spins high enough to investigate such phenomena.

The energy displacement $\delta E$ and rotational frequency ratios $\omega ^\mp/\omega ^\pm$
between the opposite parity states of the same spin, deduced from the current study are depicted in
\figurename~\ref{143La-DE-frequency-ratio}(a) and (b) respectively.
For stabilized octupole shape, the expected values are $\delta E=0$ and $\omega ^\mp/\omega ^\pm = 1$,
which are shown as dotted lines in the respective plots.
It can be seen from \figurename~\ref{143La-DE-frequency-ratio}(a) that the $\delta E$ values for
$s=\pm i$ doublets both decrease sharply with increasing spin and reach
the stable octupole limit at high spin. The variation of $\delta E$ does not show any dependence on the
simplex quantum number within the observed spin range, which is expected in the case of
stabilized octupole deformation.  
A similar decreasing trend was observed in the case of $N=86$ isotones of Ba ($^{142}$Ba) and
Ce ($^{144}$Ce)~\cite{Luo2009}. Furthermore, the $^{144}$Ce nucleus reaches the stable limit at spin 7,
while $^{142}$Ba does not reach the stable limit.

Indication about the nature of octupole excitation can also be obtained from the quantity
$\omega ^\mp/\omega ^\pm = 1$, i.e. the rotational frequency ratio of the opposite parity bands. 
As can be seen from \figurename~\ref{143La-DE-frequency-ratio}(b), the rotational frequency
ratio approaches unity (the limit corresponding to stable octupole deformation)
for higher spins for the $s=-i$ structure. The limited data for the  $s=i$ structure
did not allow us to obtain the nature of its variation with spin.  
The behaviour of the parity doublet bands
in $^{143}$La is found to be similar to that in $^{142}$Ba~\cite{Luo2010}.
Thus the various observables presented in this work, indicate the presence of stable octupole
deformation at high spins in $^{143}$La.



\section{Summary}

The high spin states in $^{140-142}$La have been reported for the first time.
The previous assignments of transitions and levels in $^{142}$La reported in literature,
are now assigned to $^{143}$La from  direct (A,Z) identification.
The present results could be obtained, thanks to two powerful complementary measurements of
high-fold gamma rays from Cf fission and  $\gamma$ rays in coincidence with  isotopically
identified fission fragments produced in fission of  $^{238}$U + $^9$Be system around the barrier.
The level structures are interpreted in terms of the systematics of neighbouring odd-Z
nuclei above $Z = 50$ shell closure and large-scale shell model calculations.
The present measurement for $^{140-143}$La isotopes  provides the missing information
for understanding the evolution of the excitation modes in the La isotopic chain above
the $N =82$ shell closure.
Based on the present work, the presence of a pair of parity doublet band structures
is established in $^{143}$La, not routinely observed in many nuclei.
The properties of these parity doublet band structures indicate the presence
of stable octupole deformation in $^{143}$La. 
Based on the known structures of the larger-N La isotopes, the octuple
deformation is observed to decrease while going towards the normally expected region of
octupole collectivity near N$\approx$90 unlike in nearby isotopes. 
Further Potential Energy Surface (PES) / Total Routhian Surface (TRS) /
Cranked Shell Model (CSM) calculations or microscopic shell model calculations
using a larger model space could give a greater insight into this  behaviour
in the La isotopes. Separation energies obtained from new precision mass
measurement~\cite{jaries2024} have shown the strongest change of two-neutron separation energy
in the periodic table between $N=92$ and $N=94$ in La. This motivates a deeper
investigation of the level structure of nuclei like $^{153}$La.

\section{Acknowledgements}

The authors gratefully acknowledge the AGATA collaboration for the availability of the AGATA
$\gamma$ tracking array at GANIL.  We would like to thank the GANIL accelerator staff for their
technical contributions. We thank A. O. Macchiavelli for help in data collection during the
experiment. SB, SB and RB acknowledge the support received from CEFIPRA project
No. 5604-4 and SB, RP acknowledge the support from the LIA France-India agreement.  PB and AM
acknowledge support from the Polish National Science Centre (NCN) under Contract No.
2016/22/M/ST2/00269 and the French LEA COPIGAL project.  HLC and PF acknowledge support from the
U.S. Department of Energy, Office of Science, Office of Nuclear Physics under Contract
No. DE-AC02-05CH11231 (LBNL). RMPV acknowledges partial support from the Ministry of Science, Spain, under
the grants BES-2012-061407, SEV-2014-0398, FPA2017-84756-C4 and from EU FEDER funds.
The work at Vanderbilt University and Lawrence Berkeley
National Laboratory are supported by the U.S. Department of
Energy under Grant No. DE-FG05-88ER40407 and Contract
No. DE-AC03-76SF00098. EW also acknowledges support from Shandong Provincial Outstanding Young Scholars Fund (Overseas) Program No. 2025HWYQ-028 and Shandong University Qilu Young Scholar.

\appendix
\setcounter{table}{0}
\section{Tables of Transition Energies and Intensities}

This is the beginning of appendix.

\begin{longtable}[h]{|c|c|ccc|c|}
\caption{\label{tab:Table1}The energies (E$_\gamma$) and relative intensities (I$_\gamma$) of the 
  $\gamma$ rays assigned to $^{140}$La isotopes along with the spin and parity
  of the initial ($J^{\pi}_i$) and the final ($J^{\pi}_f$) states and the energy
  of the initial state (E$_i$). } \\
  
\hline 
$ E_{\gamma}$(keV) & $E_i$ (keV)   & $ J^{\pi}_i $&$ \rightarrow $&$ J^{\pi}_f $ & $ I_{\gamma}(Err.) $ \\

\hline
 181 & 230&$ (7^{-})    $&$ \rightarrow $&$  6^{-}   $  &95	(10)	     \\
 211 &  260& ---    &$ \rightarrow $&$  6^{-}   $      &36	(5)	     \\
 236&  285&$ (7^{-}) $&$ \rightarrow $&$  6^{-}   $  &98	(12)	     \\
 245&  1617&$ (10^{-}) $&$ \rightarrow $&$  (9^{-}) $  &100	(11)	     \\
 416&  2033&$ (11^{-}) $&$ \rightarrow $&$ (10^{-}) $  &57	(5)	     \\
 603&  2220&$ (11^{+})  $&$ \rightarrow $&$  (10^{-})   $     &53	(6)	     \\
 724 & --- & --- & $ \rightarrow $ & --- & 34 (5) \\
 1141&  1372&$ (9^{-}) $&$ \rightarrow $&$ (7^{-}) $    &87	(10)	     \\
 1223&  1508&$ (9^{-})    $&$ \rightarrow $&$  (7^{-})$  &25	(4)	     \\
 1262&  1522& --- &$ \rightarrow $&  ---           &26	(5)	     \\
 \hline
\end{longtable}


\begin{longtable}[h]{|c|c|ccc|c|}
\caption{\label{tab:Table2}The energies (E$_\gamma$) and relative intensities (I$_\gamma$) of the $\gamma$ rays assigned to $^{141}$La isotopes along with the spin and parity of the initial ($J^{\pi}_i$) and the final ($J^{\pi}_f$) states and the energy of the initial state (E$_i$). } \\
\hline 
$ E_{\gamma}$(keV) & $E_i$ (keV) & $ J^{\pi}_i $ & $ \rightarrow $ & $ J^{\pi}_f $ & $ I_{\gamma}(Err.) $ \\
\hline

 163 & 1230&$ (15/2^{+})    $&$ \rightarrow $&$  (13/2^{+})   $  &21	(2)	     \\
 189 & ---& ---& $ \rightarrow $ &--- & 9 (1)\\
 234 &  1867&$ (19/2^{+})   $&$ \rightarrow $&$  (17/2^{+})   $  &11	(2)	     \\
 320&  2729&$ (23/2^{+}) $&$ \rightarrow $&$  (21/2^{+})      $  &4	(1)	     \\
 403&  1633&$ (17/2^{+}) $&$ \rightarrow $&$  (15/2^{+})      $  &21	(3)	     \\
 410&  2278& ---  &$ \rightarrow $&$  (19/2^{+})  $            &10	(2)	     \\
 416&  2694& --- &$ \rightarrow $& ---                      &30	(3)	     \\
 439&  1067&$ (13/2^{+})  $&$ \rightarrow $&$  (11/2^{+})   $   &19	(2)	     \\
 511&  2278& --- &$ \rightarrow $& ---              &31	(4)	     \\
 517&  1067&$ (13/2^{+})    $&$ \rightarrow $&$  (9/2^{+})   $  &26	(3)	     \\
 537&  1767& --- &$ \rightarrow $&$  (15/2^{+})   $           & --- 	\\
 549&  1767& --- &$ \rightarrow $& ---                    &27	(7) \\
 551&  551&$ (9/2^{+}) $&$ \rightarrow $&$ (7/2^{+}) $         &34       (7)    \\
 566&  1633& $(17/2^{+})$   &$ \rightarrow $&$ (13/2^{+}) $     &17	(1)	     \\
 590&  1218& --- &$ \rightarrow $&$ (11/2^{+}) $            &29	(3)	     \\
 602&  1230&$ (15/2^{+})    $&$ \rightarrow $&$  (11/2^{+}) $  &60	(6)	     \\
628&  628&$ (11/2^{+})    $&$ \rightarrow $&$  (7/2^{+})   $  &100	(11)	     \\
637&  1867&$ (19^/2{+})    $&$ \rightarrow $&$  (15/2^{+}) $  &21	(3)	     \\
776&  2409&$ (21/2^{+})    $&$ \rightarrow $&$  (17/2^{+}) $  &23	(3)	     \\
862&  2729&$ (23/2^{+})    $&$ \rightarrow $&$  (19/2^{+}) $  &14	(2)	     \\
957&  3650& --- &$ \rightarrow $& ---              &12	(2)	     \\
\hline
\end{longtable}

\begin{longtable}{|c|c|ccc|c|}
\caption{\label{tab:Table3}The energies (E$_\gamma$) and relative intensities (I$_\gamma$) of the $\gamma$ rays assigned to $^{142}$La isotopes along with the spin and parity of the initial ($J^{\pi}_i$) and the final ($J^{\pi}_f$) states and the energy of the initial state (E$_i$). } \\
\hline 
$ E_{\gamma}$(keV) & $E_i$ (keV) & $ J^{\pi}_i $ & $ \rightarrow $ & $ J^{\pi}_f $ & $ I_{\gamma}(Err.) $ \\
\hline
 82 & 326&$ (6^{-})    $&$ \rightarrow $&$  (5^{-})   $  &64	(7)	     \\
 98 &  244&$ (5^{-})   $&$ \rightarrow $&$  (4)^{-}   $   &100	(14)	     \\
 120&  446&$ (7^{-}) $&$ \rightarrow $&$  (6^{-})   $    &64	(9)	     \\
 163 & --- & --- & $ \rightarrow $ & --- & 13 (2)\\
210 & --- & --- & $ \rightarrow $ & --- & 8 (1)\\
269 & --- & --- & $ \rightarrow $ & --- & 12 (2)\\
463 & --- & --- & $ \rightarrow $ & --- & 12 (2)\\
502 & --- & --- & $ \rightarrow $ & --- & 43 (5)\\
 532&  858&$ (7^{-}) $&$ \rightarrow $&$  (6^{-})   $    &21	(3)	     \\
 550 & --- & --- & $ \rightarrow $ & --- & 23 (3)\\
 562 & --- & --- & $ \rightarrow $ & --- & 21 (3)\\
 600&  925&$ (8^{-}) $&$ \rightarrow $&$  (6^{-}) $      &50	(5)	     \\
 614&  858&$ (7^{-}) $&$ \rightarrow $&$ (5^{-}) $       &21	(3)	     \\
 622&  1068&$ (9^{-})  $&$ \rightarrow $&$  (7^{-})   $   &75	(7)	     \\
 678&  1603&$ (10^{-}) $&$ \rightarrow $&$ (8^{-}) $     &29	(3)	     \\
 684&  1752&$ (10^{-})    $&$ \rightarrow $&$  (9^{-})$  &12	(2)	     \\
 724&  1792&$ (11^{-}) $&$ \rightarrow $&$  (9^{-})   $  &36	(5)	     \\
 \hline
\end{longtable}

\begin{longtable}{|c|c|ccc|c|}

\caption{\label{tab:Table4}The energies (E$_\gamma$) and relative intensities (I$_\gamma$) of the  
  $\gamma$ rays assigned to $^{143}$La isotopes along with the spin and parity of the initial 
  ($J^{\pi}_i$) and the final ($J^{\pi}_f$) states and the energy of the initial state (E$_i$). } \\

\hline 
$ E_{\gamma}(keV)$ & $E_i (keV)$ & $ J^{\pi}_i $ & $ \rightarrow $ & $ J^{\pi}_f $ & $ I_{\gamma}(Err.) $ \\

\hline

\hline
 93.9 & 1452.8 &$ 15/2^{+}    $&$ \rightarrow $&$  (13/2^{-})   $  & <16  	       \\
(128.0) &  1452.8 &$ 15/2^{+}   $&$ \rightarrow $&$  (13/2^{+})   $  & <15           \\
133.4 &  424.7 &$ (9/2^{+}) $&$ \rightarrow $&$  7/2^{+}   $    & 50 (11) 	\\
156.0 &  1698.8 &$ (17/2^{+}) $&$ \rightarrow $&$  (15/2^{-}) $	  & 2.8 (6)  	\\
170.5 &  2373.6 & --- &$ \rightarrow $& ---		  & 7 (2)  	\\
202.6 &  3830.9 & --- &$ \rightarrow $&$ (29/2^{+}) $			  & <26    	\\
204.5 &  993.9 &$ (13/2^{+})  $&$ \rightarrow $&$  11/2^{+}   $   & 20 (4)   	\\
230.3 &  2373.6 &$ (21/2^{+}) $&$ \rightarrow $&$ (19/2^{+}) $		  & 12 (2)  	\\
261.4 &  291.2 &$ 7/2^{+}    $&$ \rightarrow $&$  3/2^{+}   $  & 194 (11)   	\\
291.2 &  291.2 &$ 7/2^{+} $&$ \rightarrow $&$  7/2^{+}   $ 	  & 1000 (52)   	\\
302.5 &  2373.3 &$ (21/2^{+}) $&$ \rightarrow $&$ (19/2^{-}) $	& 43 (6)          \\
325.7 &  1778.6 &$ (17/2^{+}) $&$ \rightarrow $&$ 15/2^{+} $	  & 241 (24)          \\
326.0 &  2798.3 &$ (23/2^{+}) $&$ \rightarrow $&$ (21/2^{-}) $		  & 4 (1)  	\\
(326.0) &  (3340.7) & --- &$ \rightarrow $&$  (25/2^{+})	$ & <10	        	\\
329.0 &  2472.3 &$ (21/2^{-})    $&$ \rightarrow $&$  (19/2^{+})   $   & 1.4 (3)  	\\
364.5 &  789.3 &$ 11^/2{+}    $&$ \rightarrow $&$  (9/2^{+})	$ & 190 (16)   	\\
(364.7) &  2143.2 &$ (19/2^{+})    $&$ \rightarrow $&$  (17/2^{+})	$ & <5        	\\
371.8 &  2070.6 &$ (19/2^{-})    $&$ \rightarrow $&$  (17/2^{+})	$ & 75 (7)   	\\
394.8 &  1937.6 &$ (17/2)    $&$ \rightarrow $&$  (15/2^{-})   $ & 25 (3)   	\\
404.7 & 1857.5 &$ (17/2^{-})    $&$ \rightarrow $&$  15/2^{+}   $  & 62 (6)   	\\
424.3 &  2203.1 & --- &$ \rightarrow $&$  (17/2^{+})   $  & 17 (4) 	\\
424.7 &  424.7 &$ (9/2^{+}) $&$ \rightarrow $&$  7/2^{+}   $    & 1300 (130) 	\\
(424.8) &  2798.3 &$ (23/2^{+}) $&$ \rightarrow $&$  (21/2^{+}) $	  & <8       	\\
435.0 &  2372.6 &$ (21/2) $&$ \rightarrow $&$  (17/2) $		  & 6 (1)  	\\
453.8 &  1778.6 &$ (17/2^{+}) $&$ \rightarrow $&$ (13/2^{+}) $	& 92 (8)   	\\
458.7 &  1452.8 &$ 15/2^{+}  $&$ \rightarrow $&$  (13/2^{+})   $   & 56 (5)   	\\
(489.0) &  (2862.6) & --- &$ \rightarrow $&$ (21/2^{+}) $		  & <26        	\\
498.1 &  789.3 &$ 11/2^{+}    $&$ \rightarrow $&$  7/2^{+}   $  & 753 (60)     	\\
498.4 &  1857.5&$ (17/2^{-}) $&$ \rightarrow $&$  (13/2^{-})   $	& 32 (2)   	\\
517.7 &  4348.6 & --- &$ \rightarrow $& ---	& 26 (6)         \\
527.8 &  2070.6 &$ (19/2^{-}) $&$ \rightarrow $&$ (15/2^{-}) $	  & 26 (2)    \\
535.6 &  1324.9 &$ (13/2^{+}) $&$ \rightarrow $&$ 11/2^{+} $ & 156 (10)   	\\
548.9 &  1542.8 &$ (15/2^{-})   $&$ \rightarrow $&$  (13/2^{+})	$ & 115 (10)   	\\
569.2 &  993.9 &$ (13/2^{+})    $&$ \rightarrow $&$  (9/2^{+})   $   & 567 (49)   	\\
569.8 &  1359.1 &$ (13^/2{-})    $&$ \rightarrow $&$  11/2^{+}	$ & 145 (10)   	\\
595.0 &  2373.6 &$ (21/2^{+})    $&$ \rightarrow $&$  (17/2^{+})	$ & 194 (20)     	\\
613.6 &  3628.3 &$ (29/2^{+})    $&$ \rightarrow $&$  (25/2^{+})	$ & 40 (10)     	\\
614.9 &  2472.3 &$ (21/2^{-})    $&$ \rightarrow $&$  (17/2^{-})   $ & 18 (4)  	\\
641.1 &  3014.7 &$ (25/2^{+})    $&$ \rightarrow $&$  (21/2^{+})   $  & 84  (9)   	\\
655.1 &   2798.3 &$ (23/2^{+}) $&$ \rightarrow $&$  (19/2^{+})   $	  & 23 (3)   	\\
663.6 &  1452.8 &$ 15/2^{+} $&$ \rightarrow $&$ 11/2^{+} $	 & 374 (21)  \\
667.9 &  2738.5 &$ (23/2^{-}) $&$ \rightarrow $&$ (19/2^{-}) $	  & 42 (6)   \\
674.6 &  2373.3 &$ (21/2^{+}) $&$ \rightarrow $&$ (17/2^{+}) $		  & 12 (2)  	\\
690.4 &  2143.2 &$ (19/2^{+})    $&$ \rightarrow $&$  15/2^{+}	$ & 88 (9)   	\\
704.9 &  1698.8 &$ (17/2^{+})    $&$ \rightarrow $&$  (13/2^{+})   $   & 178 (16)   	\\
724.1 &  3097.7 & --- &$ \rightarrow $&$  (21/2^{+})	$ & 30 (4)   	\\
750.3 &  2203.1 & --- &$ \rightarrow $& $15/2^{+}$ & 22 (3)   	\\
974.6 &  1763.9 & --- &$ \rightarrow $&$  11/2^{+}	$ & 13 (3)   	\\

\hline


\end{longtable}

\end{document}